\documentclass[a4paper,11pt,openany]{article}

\usepackage{jcappub} 
\usepackage[T1]{fontenc} 
\usepackage{graphicx}
\usepackage{txfonts}

\usepackage{amssymb}
\usepackage{amsmath}
\usepackage{graphicx}
\usepackage{natbib}
\usepackage{multirow}
\usepackage{multicol}
\usepackage{txfonts}
\usepackage{microtype}
\usepackage{color}
\usepackage[colorlinks=true,citecolor=blue,linkcolor=blue]{hyperref}
\usepackage{epstopdf}
\usepackage[flushleft]{threeparttable}
\usepackage{booktabs}
\usepackage{longtable,lscape}
\usepackage{subfigure}

\usepackage[T1]{fontenc} 
\usepackage{orcidlink}
\DeclareRobustCommand{\VAN}[3]{#2}
\let\VANthebibliography\thebibliography
\def\thebibliography{\DeclareRobustCommand{\VAN}[3]{##3}\VANthebibliography}
\usepackage{graphicx}	
\usepackage{amsmath}	
\usepackage{subcaption}

\newcommand{\slsim}{\texttt{SLSim}}

\newcommand{\FT}[1]{}

\title{\boldmath \centering \texttt{SLSim}: a strong lensing population simulation package}

\author[a, b, c, 1]{Narayan Khadka\orcidlink{https://orcid.org/0000-0001-5512-2716},\note{Corresponding author.}}
\author[a]{Simon Birrer\orcidlink{0000-0003-3195-5507},}
\author[d, e, f, g]{Henry Best\orcidlink{https://orcid.org/0009-0009-6932-6379},}
\author[a]{Paras Sharma,}
\author[h]{Katsuya T. Abe,}
\author[a, i]{Xianzhe TZ Tang\orcidlink{0009-0007-3185-7030},\note{Corresponding author.},}
\author[j]{Carly Mistick,}
\author[k]{Felipe Urcelay,}
\author[a]{Emrecan M. Sonmez,}

\author[l]{Nikki Arendse,}
\author[b, c]{Sydney Erickson,}
\author[l]{Jacob O. Hjortlund,}
\author[m]{Phil Holloway,}
\author[a]{Alan Huang,}
\author[a]{Rahul Karthik,}
\author[a]{Mia Lamontagne,}
\author[n, o]{Vibhore Negi,}
\author[p]{Justin R. Pierel,}
\author[q]{Bruno S\'anchez,}
\author[r]{Aysu Ece Saricaoglu,}
\author[s, t]{Anowar Shajib,}
\author[a]{Yixuan Shao,}
\author[b, u]{Padma Venkatraman,}
\author[r]{Bryce Wedig,}

\author[u]{Aadya Agrawal,}
\author[v]{Timo Anguita,}
\author[w, x]{Pedro Bessa,}
\author[w, x]{Clecio R. Bom,}
\author[a]{Sofia Castillo,}
\author[y]{Thomas Collett,}
\author[r]{Tansu Daylan,}
\author[b, c]{Steven Dillmann,}
\author[z]{Margherita Grespan,}
\author[aa]{Erin E. Hayes}
\author[ab]{R\'emy Joseph,}
\author[s, t]{Richard Kessler,}
\author[z]{Tian Li,}
\author[b, c]{Phil Marshall,}
\author[n]{Anupreeta More,}
\author[ac]{Veronica Motta,}
\author[u]{Gautham Narayan,}
\author[d, e, f]{Matt O'Dowd,}
\author[h]{Masamune Oguri,}
\author[ad]{Graham Smith,}
\author[m]{Aprajita Verma,}
\author[d, e, f]{Giorgos Vernardos,}
\author[]{the Strong Lensing Science Collaboration, and}
\author[]{the LSST Dark Energy Science Collaboration}
\affiliation[a]{Department of Physics and Astronomy, Stony Brook University, Stony Brook, NY 11794, USA.}
\affiliation[b]{Kavli Institute for Particle Astrophysics and Cosmology, Department of Physics, Stanford University}
\affiliation[c]{SLAC National Accelerator Laboratory, Menlo Park, CA, USA}
\affiliation[d]{The Graduate Center of the City University of New York, 365 Fifth Avenue, New York, NY 10016, USA}
\affiliation[e]{Department of Astrophysics, American Museum of Natural History, Central Park West and 79th Street, NY 10024-5192, USA}
\affiliation[f]{Department of Physics and Astronomy, Lehman College of the CUNY, Bronx, NY 10468, USA}
\affiliation[g]{Department of Theoretical Physics and Astrophysics, Faculty of Science, Masaryk University, Kotl\'arsk\'a 2, CZ-611 37 Brno, Czech Republic}
\affiliation[h]{Center for Frontier Science, Chiba University, 1-33 Yayoi-cho, Inage-ku, Chiba 263-8522, Japan}
\affiliation[i]{Department of Astronomy, Boston University, 725 Commonwealth Ave, Boston, MA 02215, USA}
\affiliation[j]{Department of Physics and Astronomy, University of Michigan, Ann Arbor}
\affiliation[k]{Institute of Astrophysics, Pontificia Universidad Cat\'olica de Chile, Santiago, Chile}
\affiliation[l]{OskarKlein Centre, Department of Physics, Stockholm University, SE-106 91 Stockholm, Sweden}
\affiliation[m]{Department of Physics, Oxford University, Keble Road, Oxford, OX1
3RH, UK}
\affiliation[n]{The Inter-University Centre for Astronomy and Astrophysics, Post Bag 4, Ganeshkhind, Pune 411007, India}
\affiliation[o]{The Kavli Institute for Astronomy and Astrophysics, Peking University}
\affiliation[p]{Space Telescope Science Institute, 3700 San Martin Drive, Baltimore, MD 21218, USA}
\affiliation[q]{Department of Physics and Astronomy, Duke University}
\affiliation[r]{Department of Physics and McDonnell Center for the Space Sciences, Washington University, St. Louis, MO 63130, USA}
\affiliation[s]{Kavli Institute for Cosmological Physics, University of Chicago}
\affiliation[t]{Department of Astronomy and Astrophysics, University of Chicago}
\affiliation[u]{Department of Astronomy, University of Illinois at Urbana-Champaign}
\affiliation[v]{Instituto de Astrof\'isica, Facultad de Ciencias Exactas, Universidad Andres Bello, Av. Fernandez Concha 700, Las Condes, Santi-
ago, Chile}
\affiliation[w]{Centro Brasileiro de Pesquisas F\'isicas, Rua Dr. Xavier Sigaud 150, 22290-180 Rio de Janeiro, RJ, Brazil}
\affiliation[x]{Centro Federal de Educa\c{c}\~{a}o Tecnol\'ogica Celso Suckow da Fonseca, Rodovia M\'arcio Covas, lote J2, quadra J - Itagua\'i, Brazil}
\affiliation[y]{Institute of Cosmology and Gravitation, University of Portsmouth, Burnaby Rd, Portsmouth PO1 3FX, UK}
\affiliation[z]{National Center for Nuclear Research, Andrzeja Soltana 7/3, PL-05-400 Otwock, Poland}
\affiliation[aa]{Institute of Astronomy and Kavli Institute for Cosmology, University of Cambridge, Madingley Road, Cambridge CB3 0HA, UK}
\affiliation[ab]{Department of Astrophysical Sciences, Princeton University}
\affiliation[ac]{Instituto de Fisica y Astronomia, Universidad de Valparaiso, Chile}
\affiliation[ad]{School of Physics and Astronomy, University of Birmingham, Birmingham B15 2TT, United Kingdom}
\emailAdd{narayan@slac.stanford.edu}
\abstract{Gravitational lensing offers unique insights into cosmology by bending light around massive objects. Strong gravitational lensing, in particular, produces magnified and often multiple images of distant sources, crucial for precise cosmological measurements and understanding the distribution of dark matter in the universe. Current studies are limited by the number of strong gravitational lenses. From upcoming cosmological surveys, we anticipate observing a several orders of magnitude increase in the number of lenses, for both static and transient phenomena. However, detecting and analyzing these events from vast surveys like Vera C. Rubin Observatory Legacy Survey of Space and Time (LSST) presents significant challenges. To prepare for these challenges, we introduce \slsim\,, a versatile simulation tool tailored for the Vera C. Rubin Observatory. \slsim\, integrates advanced astrophysical models with computational efficiency to generate synthetic strong lens populations under realistic observational conditions. \slsim\, simulates static and variable lensing scenarios, essential for cosmological studies, training and testing lens search and data analysis pipelines. This paper details \slsim\,'s design and implementation, emphasizing its modularity and capabilities across various astrophysical regimes. Validation against observational data and existing simulations confirms \slsim\,'s accuracy in reproducing observed lensing phenomena. \slsim\, is publicly available at \href{https://github.com/LSST-strong-lensing/slsim}{https://github.com/LSST-strong-lensing/slsim}, and we anticipate continued development and expansion of its capabilities. Users are encouraged to check the repository for updates and to contribute to ongoing community efforts in strong lensing simulations.}

\begin{document}
\maketitle
\section{Introduction}
\label{sec:Introduction}
Gravitational lensing, an astrophysical phenomenon, occurs when massive objects, such as galaxies or clusters of galaxies, bend the path of light from background sources. Among the various types of gravitational lensing, strong lensing is particularly significant because of its ability to create highly magnified and often multiply-imaged views of distant sources. This phenomenon not only enables a measurement of the current expansion rate of the universe ($H_0$) but also enables us to study the distribution of dark matter and dark energy in the universe \citep[e.g., ][]{Link1999, Megneghettietal2005, Julloetal2010, Collettetal2014, Wongetal2015, Boltonetal2015, Lietal2016, Kamadaetal2016, Wongetal2017, Bonvinetal2017, Gilman2018, maganaetal2018, Birrer2020, Birrer_review, Keeley2024, Tan_et_el2024, Lietal2024, Gilmanetal2024, Birreretal2024, Tokayeretal2024}.

The Rubin Observatory's Legacy Survey of Space and Time (LSST) \citep{LSST2019} is expected to uncover an unprecedented number of strong gravitational lenses. It is anticipated that there will be potentially $10^5$ galaxy-galaxy lenses \citep{Collett2015}, and over $200$ cosmologically useful lensed quasars \citep{OguriMarshall2010}. In addition, around 44 lensed supernovae are expected to be discovered per year \citep{Arendse:2023}. Among these, approximately 10 lensed supernovae per year are anticipated to be suitable for precise time-delay measurements \citep{Arendse:2023}. These strong gravitational lenses can be used for various cosmological studies. However, finding these events from observational data is a challenging task. As surveys like LSST and Euclid provide tremendous amounts of data, it is unfeasible to find all the strong lensing events by human inspection, which will limit our ability to fully exploit the scientific potential of these data. On the other hand, extracting accurate cosmological constraints from large sets of lenses requires sophisticated forward lens modeling, which is computationally expensive. Therefore, computationally more efficient techniques for these population-level analyses are needed that can enhance our ability to fully explore science cases through future survey data. In addition, strong lenses are rare and mostly identified in the large-scale structure mass distribution. Furthermore, strong lenses are most commonly found at intermediate redshifts $(z \sim 0.3$–$0.8)$, where the lensing efficiency is geometrically maximized due to the optimal alignment between the observer, lens, and background source. These selection effects can bias the strong lens population towards the higher mass, higher shear \citep{Beckeretal2016}, and a particular redshift range. For accurate population-level constraints, these selection biases need to be understood, and simulation of a realistic population of strong lenses can help to account for these selection biases.

Preparation for strong lensing science in large surveys requires both efficient lens-finding and modeling techniques. Machine learning (ML) methods have proven to be effective in identifying strong gravitational lenses in large data sets \citep{Estradaetal2007, Schaeferetal2018, Bometal2017, Metcalfetal2019, Bometal2022, Stein_2022, Rezaei_et_al2022, Bagetal2024, Moreetal2024, Lietal_lens_finding2024}, and are being applied to lens modeling, significantly reducing computational costs \citep{Hezavehetal2017, Wagner-Carenaetal2021, Faginetal2024, Jarugulaetal2024, Zhangetal2024, Ericksonetal2024}. These techniques rely heavily on high-quality training and testing datasets. Although limited real observations are available for testing, the lack of sufficient training data presents a major challenge. Simulated data sets are essential to fill this gap, but their effectiveness depends critically on the realism of simulated lens populations and observational conditions \citep{Metcalfetal2019, Bometal2022}. The Strong Lensing Simulation (\texttt{SLSim})\footnote{\href{https://github.com/LSST-strong-lensing/slsim}{https://github.com/LSST-strong-lensing/slsim}} package addresses this challenge by enabling the generation of realistic, survey-adapted training datasets and facilitating performance assessments of ML-based detection and modeling methods.
Simulating realistic strong lens populations is a complex task requiring accurate source and deflector models, computational efficiency, and attention to observational noise characteristics. A flexible and realistic simulation framework like \texttt{SLSim} is therefore essential for several key purposes: 1)Developing lens detection methods using simulated images that resemble real survey data. 2) Developing efficient lens modeling, light curve extraction, and analysis algorithms. 3)Correcting for selection effects in population-level analyses of strong lenses. 4) Providing a flexible simulation platform adaptable to various surveys and science goals.

Significant work has been done on population-level strong lensing catalog simulations \citep{OguriMarshall2010, Collett2015, Oguri2018, Lemon2023}. However, a more systematic and better-integrated software solution is still needed for pixel-level simulations. In this paper, we introduce a comprehensive and user-friendly \slsim\, package designed to generate detailed strong lens populations. \texttt{SLSim} is built with a modular architecture, where each module is developed for specific tasks and interfaces with others to perform integrated operations. The modular design ensures flexibility, allowing users to swap out modules for different lens and source classes. This flexibility gives users full control over their choices, allowing customization based on the specific requirements of different science cases. The software accounts for observational conditions and is capable of simulating strong lens populations for a given observatory, particularly for the Vera Rubin Observatory with adaptability to others, such as the \textit{Roman Space Telescope} \citep{Spergeletal2015, Piereletal2021} and \textit{Euclid} \citep{Euclid2022}.

The package includes its own image simulation module, which recognizes individual lens objects from the simulated lens population, allowing users to generate images for each lens. Additionally, \texttt{SLSim} is fully integrated with the Rubin software ecosystem through the LSST science pipeline module, enabling the injection of simulated lens images into observational data or sky simulations provided by LSST-DESC\footnote{The LSST Dark Energy Science Collaboration (DESC) is a scientific collaboration working to extract cosmological constraints from LSST data. More information is available at \href{https://lsstdesc.org/}{https://lsstdesc.org/}.}. This integration places the simulated lenses in a real astrophysical environment, with the final images containing strong lenses alongside other objects, such as galaxies or stars, resulting in highly realistic lens images. \texttt{SLSim} is capable of simulating static and variable lens systems, and all of these tasks can be performed at the population level, providing a realistic strong lens image catalog.

We begin by describing the design of our simulation package in Section ~\ref{sec:design}, detailing its core components, functionalities, and the features that distinguish it from existing tools. In Section~\ref{sec:aspects}, we describe supporting components of \slsim\,. In Section~\ref{sec:lsst_science}, we describe the LSST science pipeline module which integrates \slsim\ with the LSST software ecosystem. In Section~\ref{sec:validation}, through some tests and validation examples, we demonstrate the package's accuracy and performance for specific scenarios. In Section~\ref{sec:notebook}, we briefly discuss the notebook suite containing many working examples. In Section~\ref{sec:future}, we discuss potential applications of this package in the LSST survey and other upcoming surveys and future directions in its development. In Section~\ref{sec:conclusion}, we present our conclusion.

\begin{figure*}
    \includegraphics[width=\linewidth]{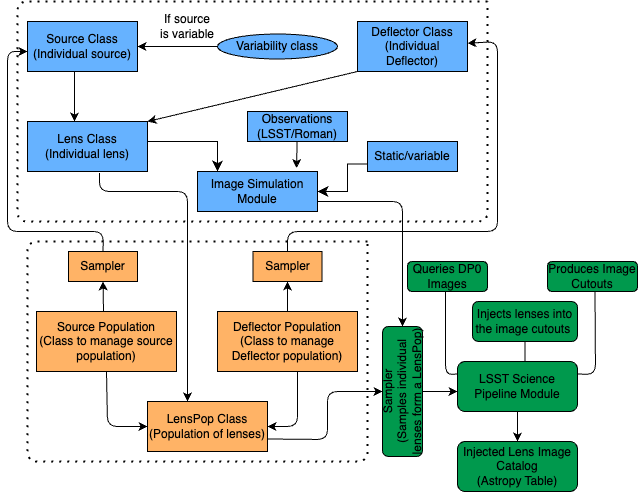}\par
\caption{Illustration of the different modules and classes of \texttt{SLSim}. In the top dotted box, all the classes and modules shown in blue handle individual objects, and these modules and classes interface with each other to produce a single lens. In the second dotted box, the classes shown in orange manage the population of corresponding objects, interfacing both with each other and with the classes involved in individual lens simulations. The blocks shown in green represent the LSST science pipeline module, along with its tasks and output. This module interfaces with the \texttt{LensPop} class and the image simulation module, generating a realistic strong lens image catalog by injecting simulated lenses into the observational data or sky simulations provided by LSST-DESC.}
\label{fig:illustration}
\end{figure*}

\section{Design}
\label{sec:design}
\slsim\, is a joint development effort between the Dark Energy Science Collaboration (DESC) and the Strong Lensing Science Collaboration (SLSC)\footnote{SLSC is dedicated to advancing strong gravitational lensing science with the LSST data. More information is available at \href{https://sites.google.com/view/lsst-stronglensing}{https://sites.google.com/view/lsst-stronglensing}.}. The software is an open-source package developed in \texttt{Python}. It follows the high standards of software development, incorporating continuous integration testing to ensure reliability and robustness across all updates. The project emphasizes accessibility, with documentation to support both developers and users. Code contributions undergo rigorous review and testing to maintain quality and consistency. This collaborative approach fosters community participation, making \slsim\, a community-driven tool for strong lensing research.

\slsim\, consists of various components, including multiple simulation modules and a module designed to integrate it into the Rubin software ecosystem. The design of the \slsim\, package is illustrated in Fig.\ \ref{fig:illustration}, with a description of this design provided below.

\subsection{Modules and classes}
\label{sec: modules}
\slsim\, is a modular software package designed to handle various aspects of simulating strong gravitational lens populations. Its structure is illustrated in Fig.\ \ref{fig:illustration}, and it primarily consists of three types of modules and classes\footnote{In the context of \slsim\, a module refers to a collection of related code, organized in a single or multiple Python files, that provides specific functionalities. A class is a Python object containing attributes and methods associated with the astrophysical object or process of interest in \slsim\,.}. The first group, shown in blue, is responsible for handling individual objects such as individual source, deflector, lens, and lens image. These classes interface with one another to simulate a single lens. The second group, depicted in orange, manages populations of these objects (source, deflector, and lens populations). These classes not only interact with each other but also interface with those involved in individual lens simulations to generate lens populations. The final component, represented in green, is the LSST science pipeline module, which integrates the package with the LSST software ecosystem. This module is designed to inject simulated lenses into the observational data or sky simulations provided by LSST-DESC. In this section, we describe the major modules and classes of \slsim\,.

\subsubsection{The \texttt{Source} class}
\label{sec: source}
The \texttt{Source} class represents a source object with the capability to handle either variable or static sources. It primarily uses a dictionary or an \texttt{astropy} table (source\_dict) to define properties such as position, ellipticity, angular size, and Sersic index. Additionally, it supports the simulation of supernova or quasar light curves through an optional variability model. When a variability model is specified, the class extracts the relevant parameters from the source dictionary and constructs a corresponding light curve using the Variability class. Depending on the nature of the source, the \texttt{Source} class uses one of three specialized classes: \texttt{ExtendedSource}, \texttt{PointSource}, and \texttt{PointPlusExtendedSource}.

Each of these subclasses handles specific source configurations. The \texttt{ExtendedSource} class employs the \texttt{SingleSersic}, \texttt{DoubleSersic}, or \texttt{Interpolated} models to represent various extended source profiles. Meanwhile, the \texttt{PointSource} class utilizes the \texttt{Quasar}, \texttt{SupernovaEvent}, and \texttt{GeneralLightCurve} classes to model point-like variable sources. The \texttt{PointPlusExtendedSource} class integrates both extended and point source components, combining the functionality of the aforementioned classes to simulate systems that contain both sources.

\subsubsection{The \texttt{Deflector} class}
\label{sec: deflector}
The \texttt{Deflector} class manages an individual deflector. The deflector can be either a galaxy or a group/cluster of galaxies. The class initializes deflector attributes such as redshift, velocity dispersion, and stellar mass based on input parameters. It offers properties like mass and light ellipticities, magnitude in specific bands, and halo characteristics. The \texttt{Deflector} class makes use of three distinct classes to handle deflector mass and light models: the \texttt{EPLSersic}, \texttt{NFWCluster}, and \texttt{NFWHernquist} class. The \texttt{EPLSersic} class combines an elliptical power-law mass profile with a Sersic light profile, making it suitable for modeling galaxies. It utilizes parameters such as velocity dispersion, mass and light ellipticities, stellar mass, and angular size to characterize the deflector's gravitational potential and photometry.

The \texttt{NFWCluster} and \texttt{NFWHernquist} classes extend the complexity of deflector systems, incorporating dark matter halos and subhalos. The \texttt{NFWCluster} class characterizes a large-scale galaxy cluster with a Navarro-Frenk-White (NFW) mass profile \citep{NFW1996} and individual subhalos, each modeled as an \texttt{EPLSersic} deflector. On the other hand, the \texttt{NFWHernquist} class combines an NFW profile for the dark matter halo with a Hernquist profile for the stellar component \citep{Hernquist1990}, providing a more detailed representation of an elliptical galaxy.

\subsubsection{The Source Population class}
\label{sec: sources}
\slsim\, has three different classes to manage the source population sampled within a certain sky area: the \texttt{Galaxies}, \texttt{PointSources}, and \texttt{PointPlusExtendedSources} class. The \texttt{Galaxies} class is designed to manage galaxy catalogs in different formats, such as an astropy table or a list of tables. It accepts parameters including cosmology and the sky area over which galaxies are sampled. The class can handle both single and double sersic profiles, which define the light distribution of galaxies, and ensure that missing parameters like ellipticities or sersic indices are assigned reasonable default values. The class also performs selection cuts through the \texttt{object\_cut()} function, ensuring that galaxies meet the specified selection criteria. Through the \texttt{draw\_source()} method, the class randomly samples a galaxy from the catalog, allowing for stochastic selection and providing unlensed galaxy properties for use in lensing calculation.

Similarly, the \texttt{PointSources} class manages point source catalog such as quasars or supernovae. It handles catalogs in astropy table format, applies parameter cuts via \texttt{kwargs\_cut}, and selects point sources that meet the criteria. The \texttt{PointPlusExtendedSources} class, which combines the functionality of both \texttt{Galaxies} and \texttt{PointSources} class, manages the catalog of point sources and their host galaxies. The class also performs selection cuts on a given catalog through \texttt{kwargs\_cut}. In both of these classes, \texttt{draw\_source()} method randomly samples a source from the catalog, allowing for stochastic selection and providing unlensed source properties for use in lensing calculation.

\subsubsection{The Deflector Population class}
\label{sec: deflectors}
\slsim\, has four distinct classes to handle various deflector populations sampled in a certain sky area: the \texttt{AllLensGalaxies}, \texttt{EllipticalLensGalaxies}, \texttt{CompoundLensHalosGalaxies}, and \texttt{ClusterDeflectors}. The \texttt{AllLensGalaxies} class is designed to manage a combined catalog of elliptical and spiral galaxies while the \texttt{EllipticalLensGalaxies} class manages only an elliptical galaxy catalog. Both classes handles missing attributes in the catalog such as velocity dispersion, ellipticities, and sersic indices. These classes also performs selection cuts on deflector catalog through the \texttt{kwargs\_cuts}. In both classes, velocity dispersions for provided deflectors are estimated using the SDSS velocity dispersion function \citep{bernadietal2010}.

The \texttt{CompoundLensHalosGalaxies} class manages the deflector population generated by \texttt{SL-Hammocks} pipeline (see Section \ref{sec:hamock}). This catalog includes both halo and galaxy parameters. This class also performs selection cuts on the deflector catalog with the provided criteria. Missing attributes are addressed using default values and compute properties such as velocity dispersion using a composite model i.e. luminosity weighted velocity dispersion for a stellar Hernquist profile and a NFW halo profile.

The \texttt{ClusterDeflectors} class manages and models clusters of galaxies, where the dark matter halo follows an elliptical NFW profile, and the individual cluster members (subhalos) are modeled using an Elliptical Power Law (EPL) profile. It takes in catalogs of galaxy clusters and their member galaxies, such as those provided by the redMaPPer algorithm \cite{Rykoff2014}, and incorporates various parameters like redshift, richness, halo mass, and concentration, applying cuts based on provided criteria. In cases where the halo mass is not specified in the cluster catalog, it is estimated from the mass-richness relation in \cite{Simet2017} or \cite{Abdullah2023}, accounting for scatter. 
Similarly, concentration is derived from the mass–concentration–redshift relation in \cite{Diemer2015}, with updated parameters from \cite{Diemer2019} using the \texttt{Colossus} package \cite{Diemer2018}. 
The class computes the mean velocity dispersions for clusters following \cite{Lokas2001}, using the characteristic radius of the NFW profile as the aperture, while the velocity dispersion for member galaxies is determined using the SDSS velocity dispersion function.

\subsubsection{The \texttt{Lens} class}
\label{sec:Lens}
The \texttt{Lens} class simulates an individual strong lens using the provided individual source and deflector. A user can pass an individual source and deflector directly to this class. On the other hand, the sampler can sample a source and a deflector from the source and deflector population and pass that pair of source and deflector to the \texttt{Lens} class. With all the information required to solve the lens equation, the \texttt{Lens} class solves the Lens equation and obtains the lensing configuration. This lensing configuration can be characterized by the source and deflector redshifts, the number of lensed images, image separations, the transverse distance between the source position and the deflector center, and the brightness of the lensed extended arc if present. The Lens class computes all these quantities for a pair of source and deflector and tests them with certain criteria to check if the produced lens configuration is detectable. If the lensing configuration passes all the criteria, it is considered a detectable strong lens, and this Lens class contains all the necessary information of the lens configuration which can be accessed as different attributes.

The Lens class interfaces with both the source and deflector classes, enabling access to their properties. Consequently, it can retrieve and manage all intrinsic source and deflector properties. The Lens class then computes the corresponding lensed quantities of a source, storing these computed values as attributes within the class. The light and mass models of a deflector and a source are stored in a dictionary along with the light profile of a source. These models are the necessary ingredients for simulating the image of a lens configuration.
\begin{figure*}
    \includegraphics[width=\linewidth]{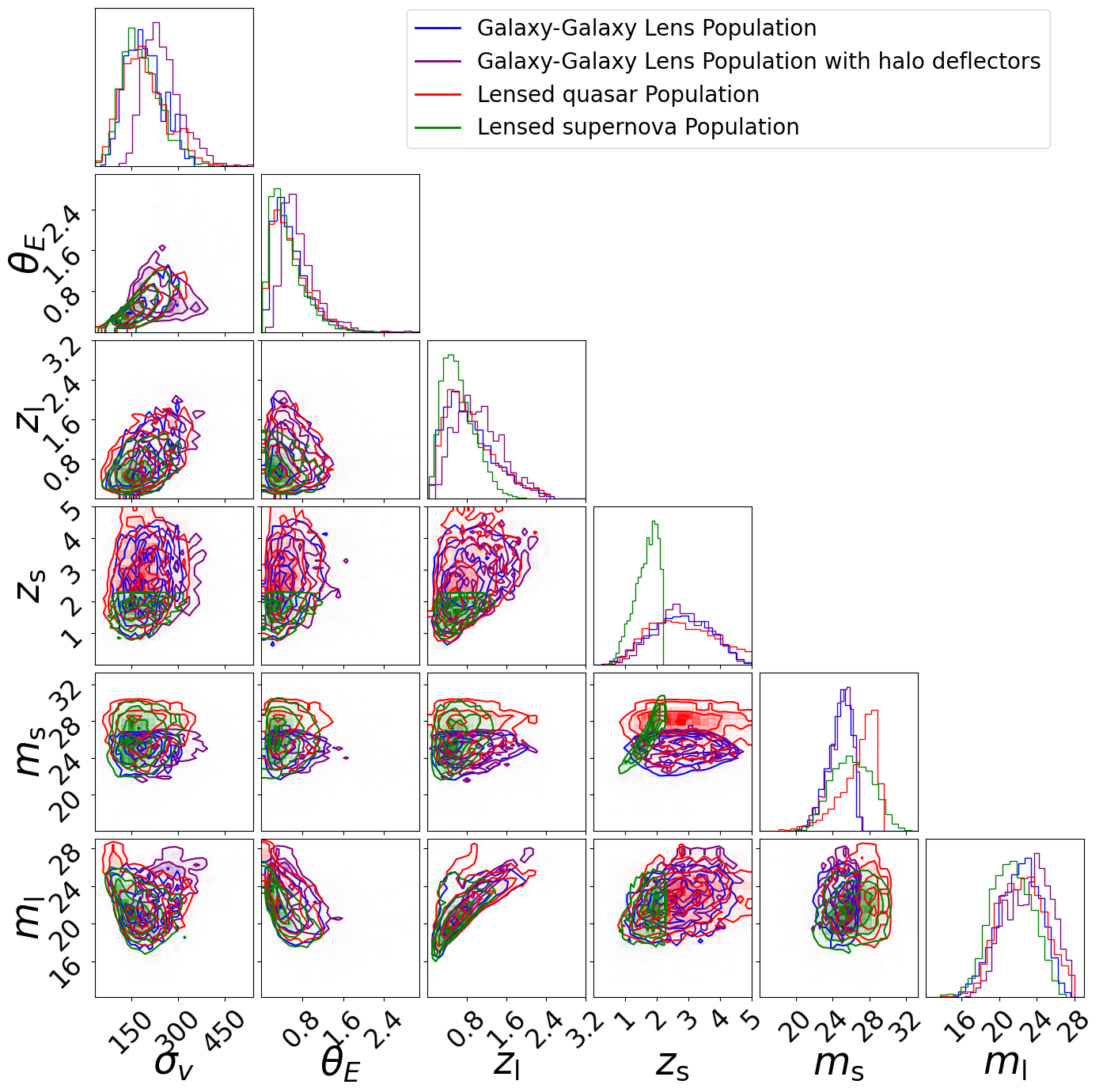}\par
\caption{Lens population of different categories simulated using \slsim\,. Galaxy-galaxy lens population with elliptical galaxies as deflectors within a 20 $deg^2$ sky area is shown in blue color. Galaxy-galaxy lens population with halo deflectors within a 10 $deg^2$ sky area is shown in purple color. Lensed quasar population within a 500 $deg^2$ sky area is shown in red color. Lensed supernovae population within a 20,000$deg^2$ sky area is shown in green color. In these plots, $\sigma_v$, $\theta_E$, $z_l$, and $m_l$ represent the velocity dispersion, Einstein radius, redshift, and i-band magnitude of the deflector galaxies, respectively, while $z_s$ and $m_s$ denote the redshift and i-band magnitude of the lensed source.}
\label{fig:galaxy_galaxy}
\end{figure*}
\subsubsection{The \texttt{LensPop} class}
\label{sec:LensPop}
This is the last layer of the code in the simulation of lens populations. The \texttt{LensPop} class deals with the population of lenses. The fundamental inputs for this class are the source population, deflector population, cosmology, and the sky area. For variable sources, one also needs to define a variability model, the time duration of lightcurve, and the corresponding parameters for these sources. The source and deflector populations are passed through the corresponding source population and deflector population classes.

\begin{figure*}
\begin{multicols}{3}
    \includegraphics[width=\linewidth]{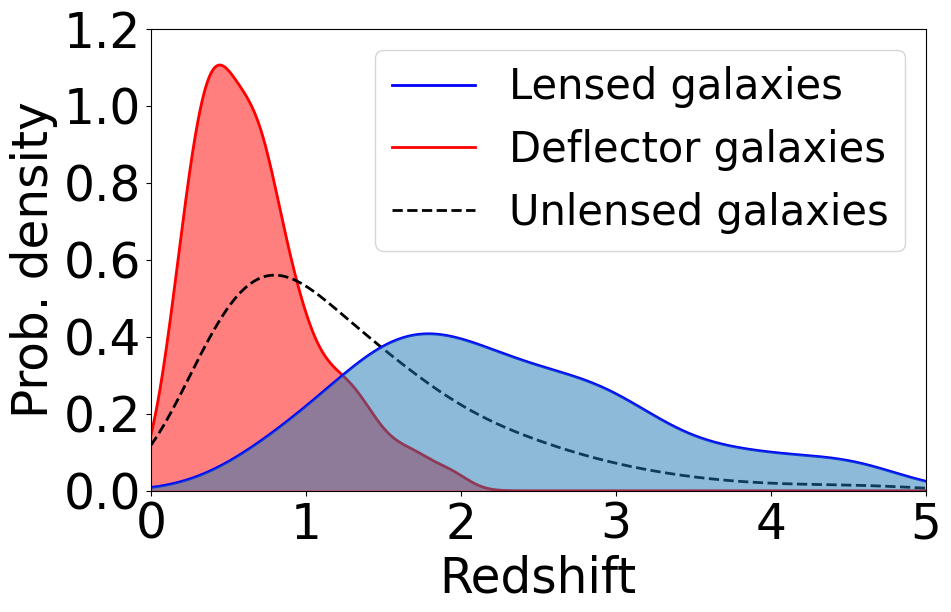}\par
    \includegraphics[width=\linewidth]{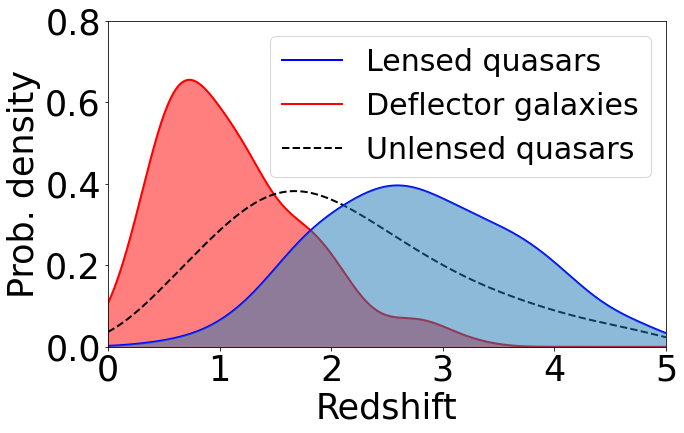}\par
    \includegraphics[width=\linewidth]{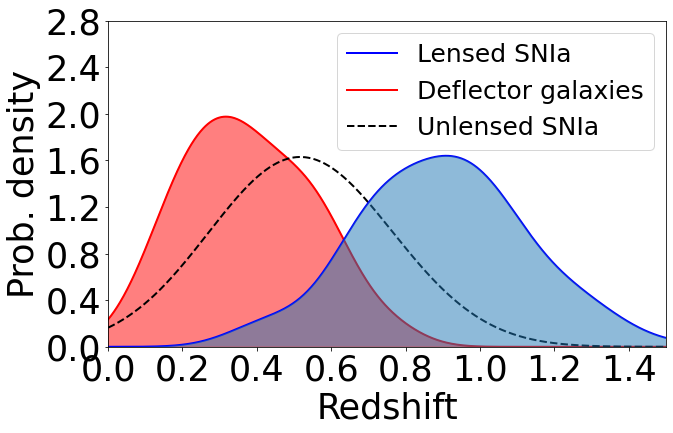}\par
\end{multicols}
\caption{Redshift distribution of deflectors (red), lensed sources (blue), and unlensed sources (black dotted) in different simulated lens populations. Left panel: Redshift distributions of deflectors, lensed galaxies, and unlensed galaxies in galaxy-galaxy lens population. For these distributions, lensed source galaxies with $i$-band magnitude less than 27 (5 year of coadd image depth) are used. Middle panel: Redshift distribution of deflectors, lensed quasars, and unlensed quasars in the lensed quasar population. For these distributions, quasars with lensed $i$-band magnitude less than 23.3 (single visit magnitude limit) are used. Right panel: Redshift distribution of deflectors, lensed supernovae, and unlensed supernovae in the lensed supernovae population. For these distributions, supernovae with lensed $i$-band magnitude less than 23.3 (single visit magnitude limit) are used.}
\label{fig:redshift_dist}
\end{figure*}

Strong lenses are rare, making the simulation of the strong lens population challenging. Identifying a lensing configuration from a vast pool of potential deflectors and sources involves substantial trial and error, as only a small fraction of configurations will result in detectable strong lensing. So, trying all possible deflector-source pairs would be computationally very expensive. To overcome this, we focus on designing simulations that optimize the selection process. For each deflector sampled from the deflector population, the \texttt{LensPop} class draws a test area around the deflector based on the deflector's Einstein radius. Then, with the source number density, the number of sources per unit sky area, computes the average number of sources to be tested in the test area. If the number of sources exceeds zero, the function iterates over the possible source-deflector configurations and initializes \texttt{Lens} class. A validity test is applied to each simulated lens system based on the specified lensing conditions, and if a system passes the test, the corresponding \texttt{Lens} class is added to the lens population. This approach efficiently narrows down the potential strong lens systems to be tested. The deflector sampling is repeated a number of times equal to the size of the deflector population, with potential source–deflector configurations tested for each sampled deflector, thereby generating a statistically representative lens sample over the specified sky area. The population of galaxy-galaxy strong lenses within a 1 $deg^2$ sky area and the population of lensed quasars and lensed supernovae within a 100 $deg^2$ are shown in Fig.\ \ref{fig:galaxy_galaxy}. Similarly, the comparison of deflector, lensed source, and unlensed source redshift distributions for each case is shown in Fig. \ref{fig:redshift_dist}.

\subsubsection{The image simulation module}
\label{sec:imsim}
Simulation of the lens population is a catalog level simulation. For many applications, a requirement is to actually generate images of lenses. The \slsim\, has an image simulation module that takes a \texttt{Lens} class instance and generates an image of the lens. This module includes various functions; some of them simulate images of static lenses, while others are tailored to simulate images of variable point sources. All these functions are encapsulated within the \texttt{lens\_image()} or \texttt{lens\_image\_series()} functions, simplifying user interaction by requiring only a Lens class instance without needing detailed knowledge about the type of deflector and source. These functions automatically identify the source category and simulate the corresponding lens images. Thus, for users, using these two functions is sufficient to simulate lens images. The \texttt{lens\_image()} function generates a single image, whereas \texttt{lens\_image\_series()} produces a time series of images for a variable lens.

\begin{figure*}
    \includegraphics[width=\linewidth]{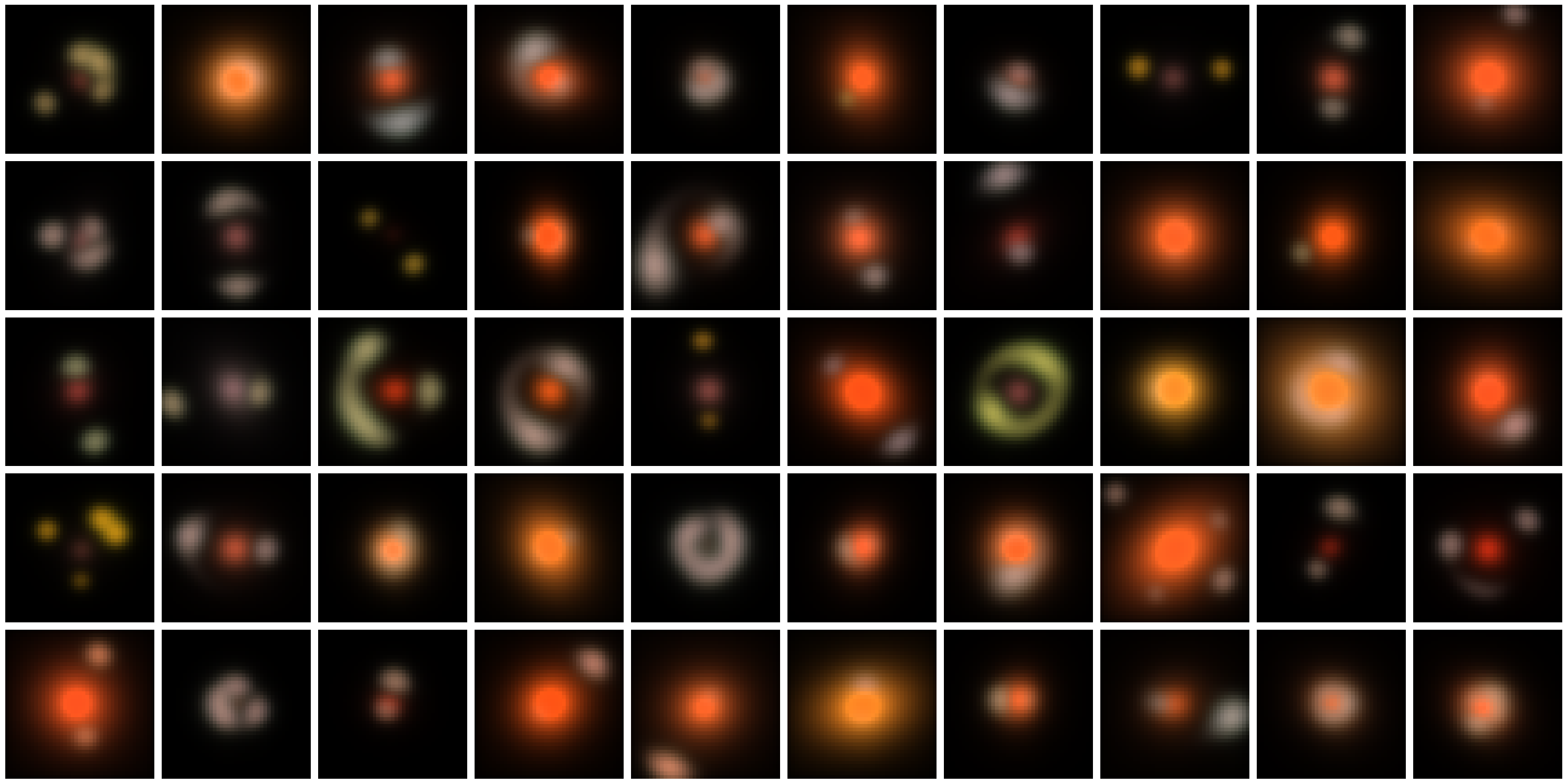}\par
\caption{Sample of galaxy-galaxy lens images simulated using \slsim\, image simulation module. These lens sample are the small subset of the lens population simulated within a $1deg^2$ sky area using \slsim\,. RGB color images are simulated using the r, i, and g band images.
}
\label{fig:galaxy_image_sample}
\end{figure*}

\begin{figure}
    \includegraphics[width=\linewidth]{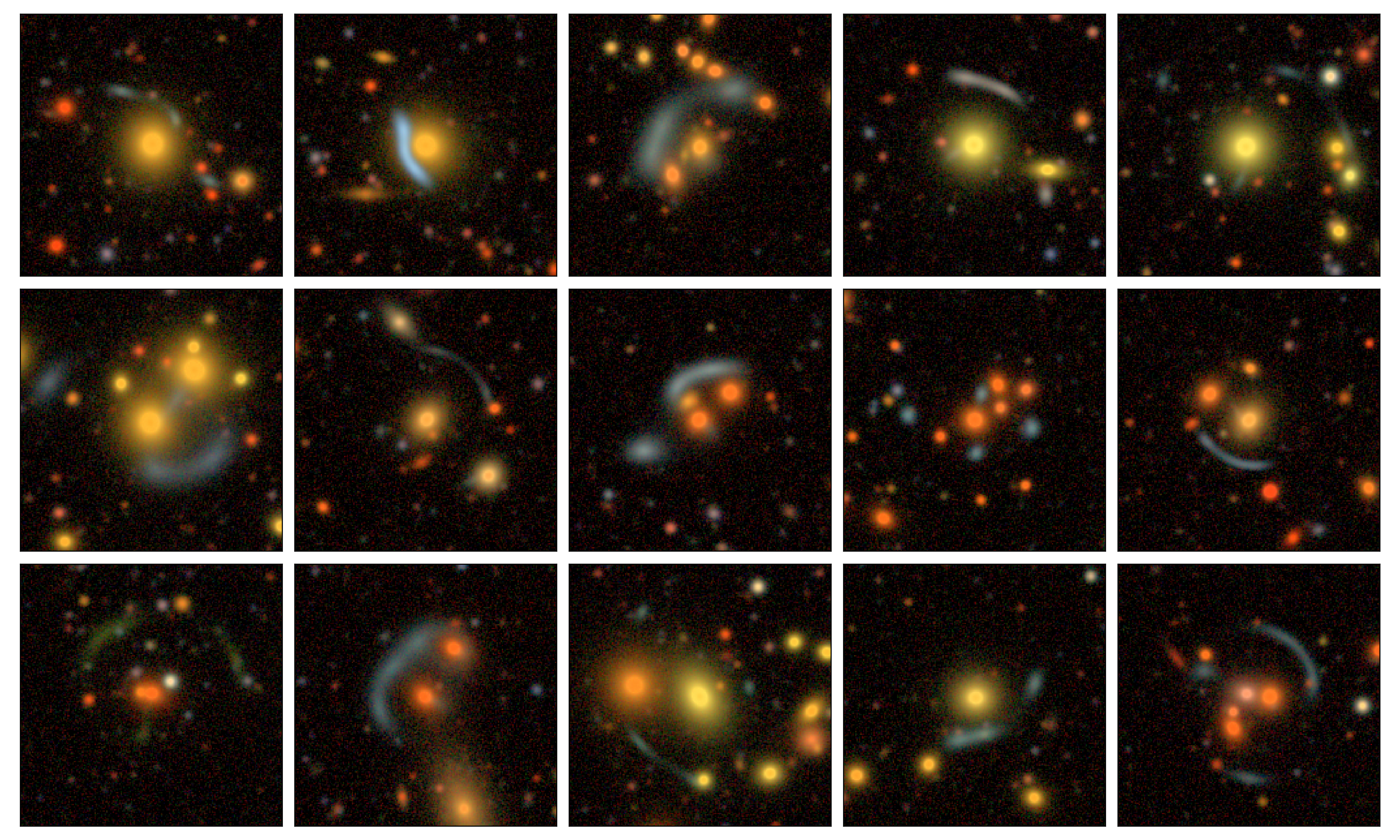}\par
\caption{Examples of cluster lenses simulated using \slsim\,. These cluster lenses are simulated using DC2 redmapper cluster catalog and injected to DP0 cutout images. RGB color images are simulated using the r, i, and g band images.}
\label{fig:clusterlens}
\end{figure}

The primary inputs for functions in the image simulation module include the \texttt{Lens} class instance, which contains comprehensive information about the lens system. Additionally, these functions accept inputs that account for observational aspects such as imaging band, zero-point magnitude of observation, point spread function (PSF) kernel, pixel-to-coordinate transformation matrix, background noise (if necessary), and observational time for variable lenses. Users can also specify the image size in terms of pixels to obtain a lens image of the desired resolution. Sample galaxy-galaxy and cluster lens images simulated using the \slsim\, image simulation module are illustrated in Figs.\ \ref{fig:galaxy_image_sample} and \ref{fig:clusterlens} respectively.

The image simulation module enables an image-based signal-to-noise ratio (SNR) calculation. SNR is a common metric for detectability in strong lens yield simulation pipelines \citep[e.g.,][]{Collett2015, Holloway2023, Ferrami2024, Cao2024}. Additionally, an image-based SNR can be used to define a mask that better isolates the lensed source than a simple annular mask. Different SNR masks could serve to model systems with different mass-to-light ratios, and to distinguish the effects of subhalos of the deflector and halos along the line-of-sight \citep[e.g.,][]{Dhanasingham2023}. The SNR calculation is a region-based algorithm established by \cite{Holloway2023} and implemented in \cite{Wedigetal2025}. A noiseless lensed source and a complete composite image (source, deflector, and background noise) are simulated to compute a per-pixel SNR, taken as the ratio of isolated source counts to total composite noise. Pixels exceeding a threshold of unity are grouped into contiguous, cross-connected regions. A cumulative SNR is then calculated for each region, yielding the maximum regional SNR as the SNR of the system.

\section{Supporting Components}
\label{sec:aspects}
In Section \ref{sec:design}, we described the major modules and classes used to simulate a realistic strong lens image catalog, which serve as the foundational framework of \slsim\,. The core inputs for the LensPop class are the source and deflector catalogs. These catalogs can be generated using different modules and pipelines within \slsim\,. For variable sources, light curves are required to simulate time-series images. We have adapted and implemented various techniques for obtaining light curves for different types of variable sources. Thus, numerous supporting codes and aspects are integrated within the foundational framework of \slsim\, to enable lens simulation. Additionally, some external packages are used to perform various tasks in the simulation of the lens population. We have described the supporting modules and aspects of \slsim\, and listed the external packages used for various tasks below.

\subsection{The \texttt{SkyPyPipeline} class}
\label{sec:skypipeline}
\begin{figure}
    \includegraphics[width=\linewidth]{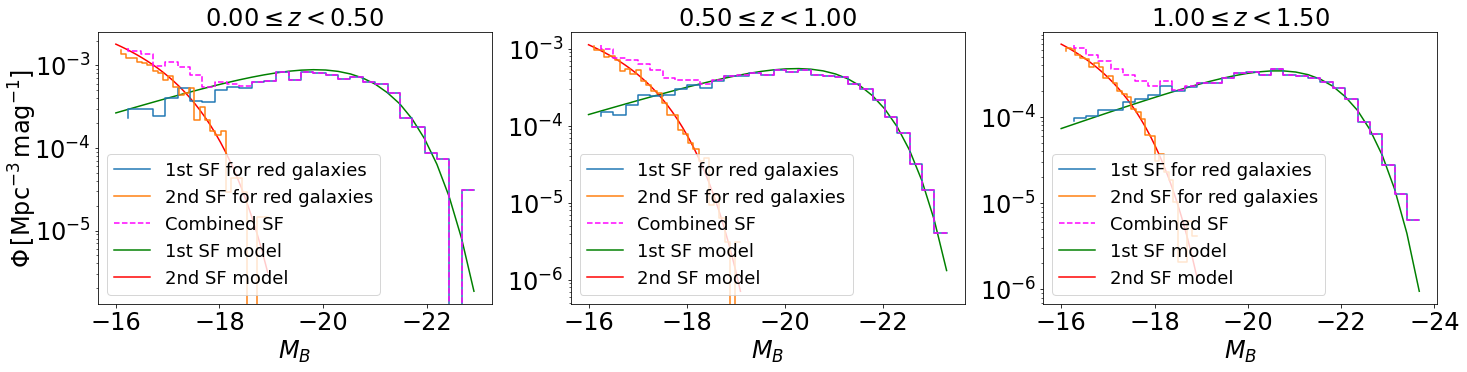}\par
    \hspace{2cm}
\caption{Comparison between the luminosity function of the elliptical galaxy population generated by \slsim\, and corresponding model predictions. The x-axis label ($M_B$) represents the absolute magnitude of a galaxy. In each plot, SF stands for the Schechter function, the histograms represent \slsim\, generated samples, and the solid curves show the model predictions.
} 
\label{fig:SF_comp}
\end{figure}

In \slsim\,, the \texttt{SkyPyPipeline} class is designed to integrate and execute simulation of a galaxy population using the \texttt{skypy} package \citep{skypyetal2021}. The user can simulate source galaxy catalogs and deflector galaxy catalogs as required. This class allows users to specify key cosmological and observational parameters, such as the cosmology, sky area, and filters, either through predefined configuration files or by modifying these parameters dynamically. If no custom configuration is provided, the class defaults to a LSST-like survey setup. The \texttt{skypy} package generates galaxy catalogs within the specified sky area, redshift range, and magnitude limits by sampling galaxies from the Schechter function \citep{Schechter1976}. For spiral galaxies, a single Schechter function is used with parameters including characteristic mass, normalization, and power law index specific to spiral galaxies. For elliptical galaxies, double Schechter functions are employed to encompass both massive elliptical and dwarf elliptical galaxies in our population. After obtaining galaxy catalogs from \texttt{skypy}, these catalogs can be passed to the source population class and the deflector population class where \slsim\,'s velocity dispersion module is called to perform SDSS velocity dispersion $(\sigma_v)$ abundance matching to ensure our galaxy populations are consistent with the observed $\sigma_v$ distribution. The comparison between the theoretical and simulated elliptical galaxy luminosity distribution is shown in Fig.\ \ref{fig:SF_comp}. An example of a velocity dispersion distribution of galaxies generated using the \texttt{skypy} pipeline within \slsim\, is shown in Fig.\ \ref{fig:galaxy_dist} (gray). These velocity dispersions are drawn from the SDSS velocity dispersion function. This is one of the fundamental parts of the simulation. This is designed to be flexible with respect to telescope filter systems; users can easily swap in custom filter transmission curves to simulate observations from other telescopes without modifying the core simulation logic.

\subsection{The \texttt{SLHammocksPipeline} class}
\label{sec:hamock}
\begin{figure}
    \centering
    \includegraphics[width=0.8\linewidth]{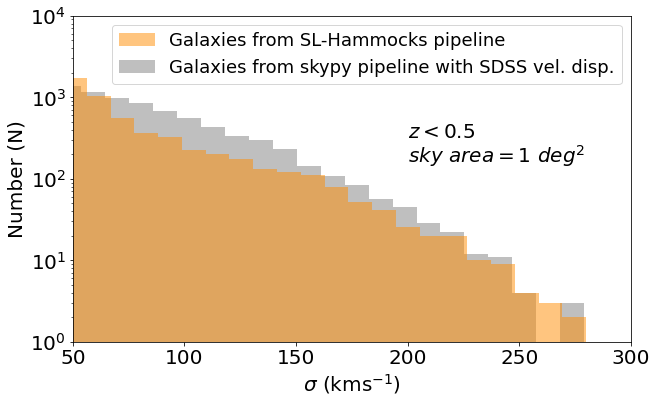}\par
\caption{
Distribution of velocity dispersion of galaxies generated using skypy and SL-Hammocks pipeline. The SDSS velocity dispersion function is used for galaxies generated using the skypy pipeline. These velocity dispersion distributions are for galaxies at $z < 0.5$ within a 1 $deg^2$ sky area.
}
\label{fig:galaxy_dist}
\end{figure}

The \texttt{SLHammocksPipeline} class is designed to simulate a deflector galaxy population using a halo-based deflector model. The halo-based deflector model samples halos from the halo mass function and populates galaxies within these halos. The implementation of this model is detailed in \citep{Katsuyaetal2025}. For more information on this model, please refer to the paper. The \texttt{SLHammocksPipeline} class allows flexibility in specifying cosmology and the sky area. This class accepts a deflector catalog from a pre-prepared CSV file. If a pre-prepared deflector catalog is not provided, the catalog is generated internally. It follows a halo model prescription, in which halo and galaxy populations are constructed based on specified redshift ranges, halo mass limits, galaxy size models, and intrinsic scatter parameters for the mass-concentration and stellar-mass-halo-mass relations. The velocity dispersions of these deflectors are computed using the velocity dispersion module, where velocity dispersions from a composite model exist for this purpose. The velocity dispersion distribution of galaxies generated using the \texttt{SLHammocksPipeline} class is shown in Fig.\ \ref{fig:galaxy_dist}.

\subsection{Point source variability}
\label{sec:variability}
For simulating time series images of variable lenses, light curves are required. Currently, SLSim handles two types of variable sources: supernovae and quasars. Supernova light curves are generated using \texttt{sncosmo} models, with \texttt{SALT2} as the default. Users can select any other light curve model supported by \texttt{sncosmo} if desired. The intrinsic light curve of each lensed supernova in the observer-frame is extracted in a specified band. These light curves are interpolated and stored in the Variability class. Subsequently, this light curve can be accessed in the Lens class, where strong lensing effects are applied to calculate the light curve for each image of the corresponding supernova.

\begin{figure*}
    \includegraphics[width=\linewidth]{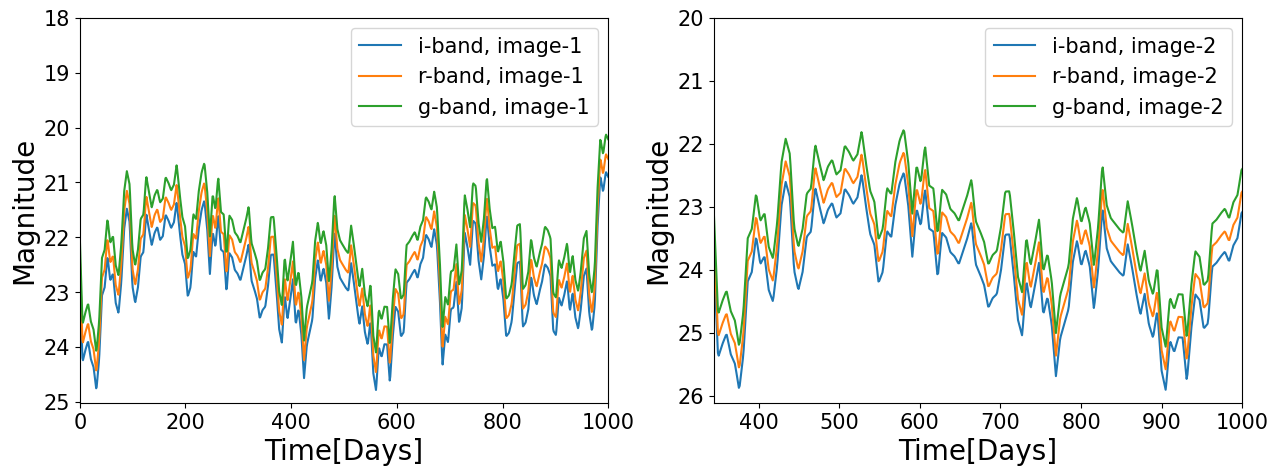}\par
    \includegraphics[width=\linewidth]{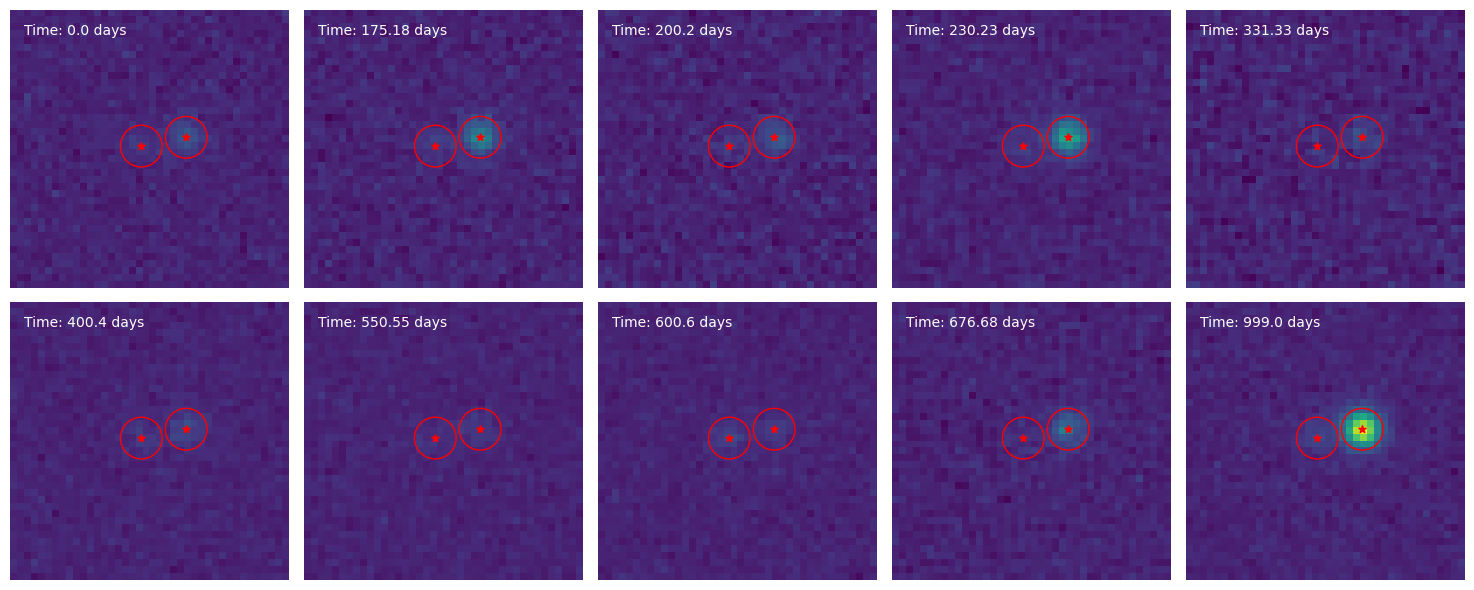}\par
\caption{Top panel: quasar lightcurves of two images of a lensed quasar generated using the quasar variability model in \slsim\,. The detailed process of generating lightcurves is described in a \href{https://github.com/LSST-strong-lensing/slsim/blob/main/notebooks/agn_intrinsic_variability_tutorial.ipynb}{quasar\_variability\_tutorial} notebook. Bottom panel: Time-series images of a lensed quasar (corresponding to the lightcurves shown in left panel) injected into the DP0 time-series single exposure images. In each sub-panel, red stars indicate the center of the corresponding images.}
\label{fig:quasar_lc}
\end{figure*}

The thin disk and lamppost models are used in SLSim for quasar variability. This model can incorporate special relativistic (Doppler beaming) and general relativistic effects (relativistic boosting and light bending) in a Kerr black hole geometry. The complete model is described in \cite{amoeba24}. SLSim has a native and efficient version of this implementation, where quasar light curves are generated assuming that they result from the accretion disk reprocessing a driving light curve. In the lamppost model, the driving light curve gets convolved with transfer functions which encode the geometry and physics of the accretion disk model.
The driving light curve is calculated using the method of \cite{Timmer95} where any power spectrum may be converted into a realization of a stochastic signal. This driving light curve is stored in the Agn class and used for all transfer functions representing any LSST band.
The transfer functions are calculated following \cite{Cackett07}, where the entire accretion disk is assumed to reverberate for every wavelength. Each LSST filter is integrated over using their discrete representation in speclite \citep{speclite23}, while time lags are calculated considering the geometry of the accretion disk including its inclination. SLSim also allows for external transfer functions to be used in place of those innately calculated to incorporate alternate reverberation models. The interpolated light curves are then stored in the Variability class and can be accessed in the Lens class, where lensing effects are applied. An example set of lightcurves in different LSST bands generated for a lensed quasar are shown in the left panel of Fig.\ \ref{fig:quasar_lc}.

\begin{figure*}
\begin{multicols}{2}
    \includegraphics[width=\linewidth]{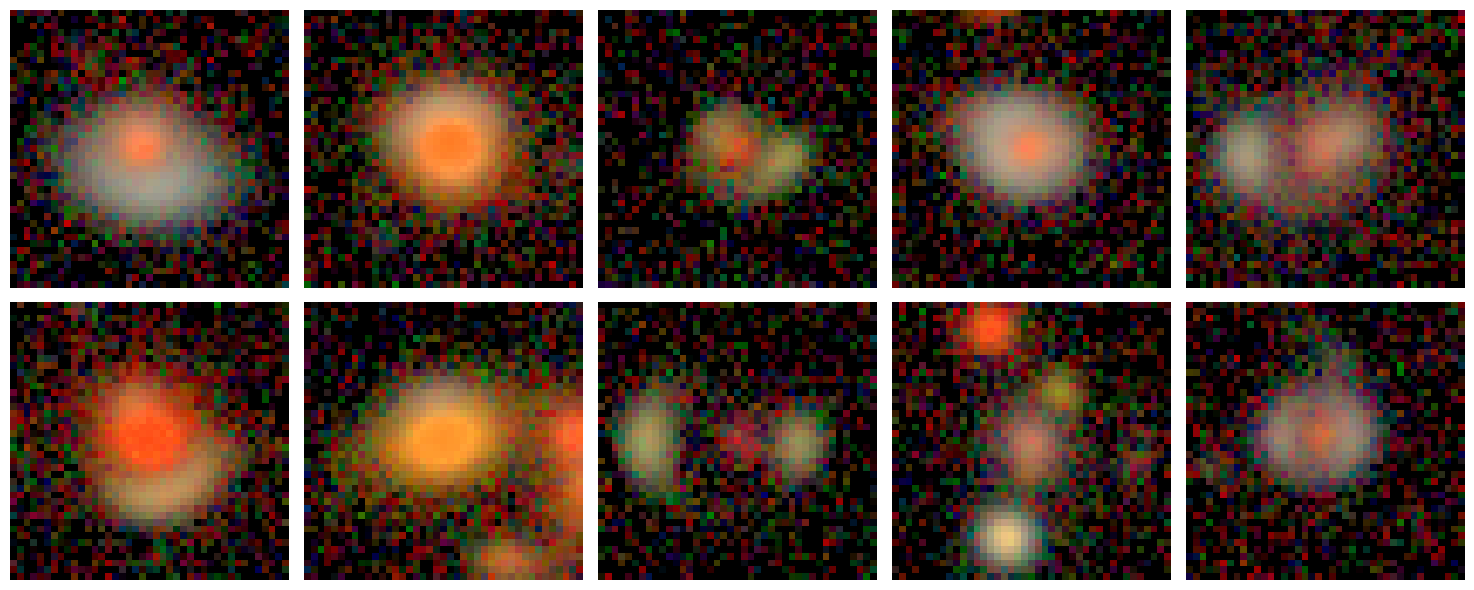}\par
    \includegraphics[width=\linewidth]{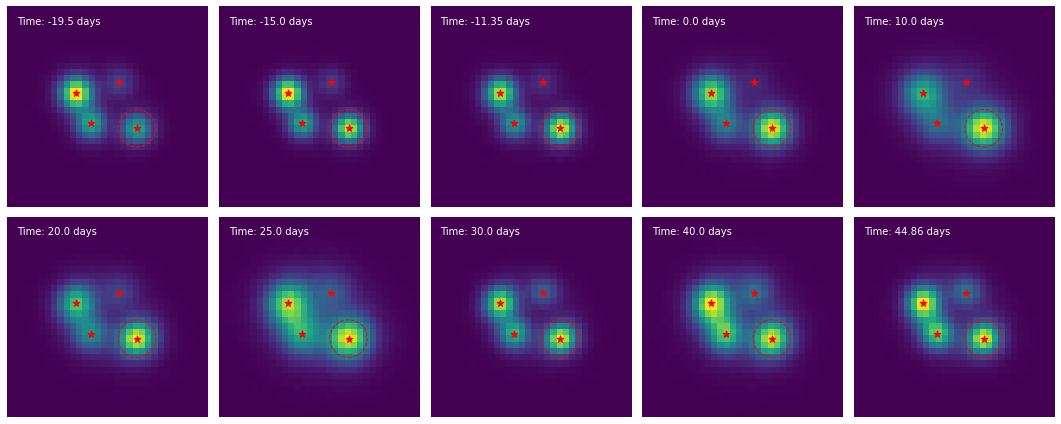}\par
\end{multicols}
\caption{Example of injected lenses using the LSST science pipeline module. Left panel: Sample of galaxy-galaxy lenses injected into the DP0 cutouts. These lens samples are a small subset of the lens population simulated within a 1 $deg^2$
sky area using \slsim\, and then injected into the DP0 cutouts. Right panel: Time-series images of a lensed supernova injected into the DP0 time-series single exposure images. In each sub-panel, red stars indicate the center of the corresponding images, and a red dotted circle indicates the first arriving image.
}
\label{fig:injected_sample}
\end{figure*}

\begin{figure}
    \includegraphics[width=\linewidth]{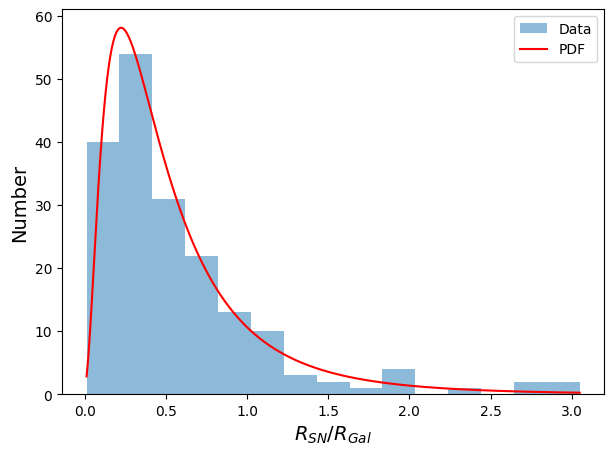}\par
\caption{Distribution of empirical supernova offset ratios within their host galaxies \citep{Wangetal2013} plotted with the obtained bestfit log-normal probability density function, with shape = 0.764609, location = -0.0284546, and scale = 0.450885.}
\label{fig:offset_ratio_dist}
\end{figure}

\begin{figure}
    \centering
    \includegraphics[width=0.9\linewidth]{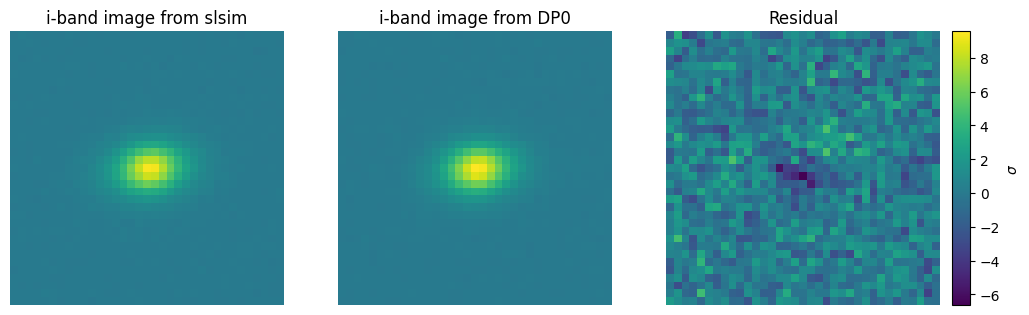}\par
    \includegraphics[width=0.9\linewidth]{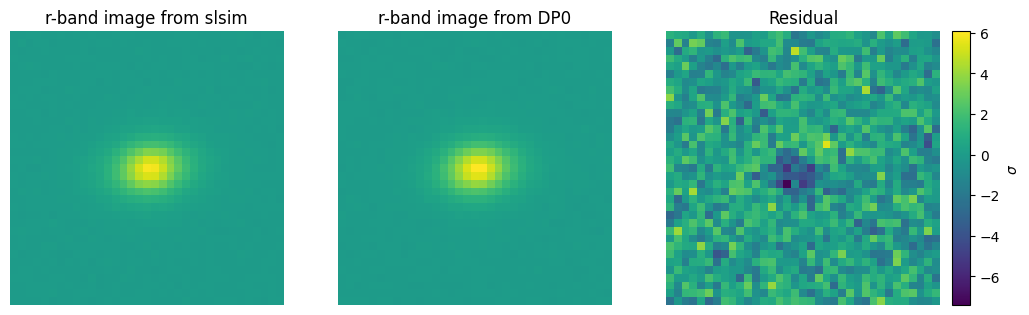}\par
    \includegraphics[width=0.9\linewidth]{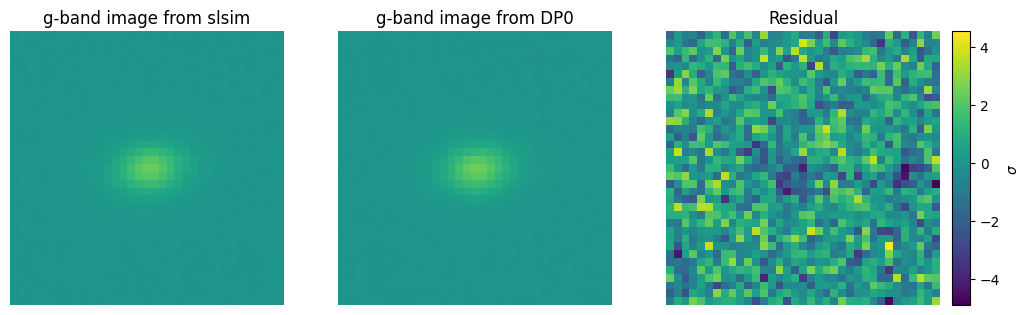}\par
\caption{Comparison of galaxy images simulated using \slsim\, in different imaging bands with the corresponding galaxy image from the DP0.2 data in the same bands. In the DP0.2 data, this galaxy is located at RA = 61.993, DEC = -36.990. The left column shows \slsim-simulated images of the galaxy in the $i$, $r$, and $g$ bands. The middle column presents the corresponding images from DP0.2 in the same bands. The right column displays the residuals between the \slsim-simulated and DP0.2 images. The residuals are calculated as (Image from \slsim\, - Image from DP0.2)/$\sigma$, where $\sigma$ represents the uncertainty in each pixel of the DP0.2 image. Each image has dimensions of $64 \times 64$ pixels.
} 
\label{fig:image_validation}
\end{figure}

\subsection{Quasar sample}
\label{sec:quasar}
For the simulation of a lensed quasar population, we need a realistic quasar population. Therefore, quasar samples need to be sampled from the observed luminosity function. We use a double power-law for the quasar luminosity function whose parameters are determined using observational data. The parametric form of this function is,
\begin{equation}
\label{eq:qlf}
    \frac{d\Phi_{\rm QSO}}{dM} = \frac{\Phi_{*}}{10^{0.4(\alpha + 1)(M - M_{*})} + 10^{0.4(\beta + 1)(M - M_{*})
}},
\end{equation}
where, $M$ is the i-band absolute magnitude of quasars. We use $\Phi^* = 5.34 \times 10^{-6} h^3 \, \text{Mpc}^{-3}$
, $\beta = -1.45$, $\alpha = -3.31$ for $z \leq 3$ and $-2.58$ for $z > 3$, and these values are obtained from the observational data \citep{Fanetal2001, Richardsetal2005, OguriMarshall2010}. $M_{*}$ is the break absolute magnitude and is given by,
\begin{equation}
\label{eq:bam}
    M_{*} = -20.90 + 5 \log h - 2.5 \log f(z),
\end{equation}
where,
\begin{equation}
\label{eq:fz}
    f(z) = \frac{e^{\zeta z} (1 + e^{\xi z_{*}})
}{\left(\sqrt{e^{\xi z}} + \sqrt{e^{\xi z_{*}}}\right)^2
}.
\end{equation}
Here $\zeta, \xi$, and $z_{*}$ are equal to 2.98, 4.05, and 1.60 respectively \citep{OguriMarshall2010}. We sample quasars from the eq.\ \ref{eq:qlf} along with eqs.\ \ref{eq:bam} and \ref{eq:fz}. The comparison of luminosity function of simulated quasar samples and prediction from eq.\ \ref{eq:qlf} in different redshift bins are shown in Fig.\ \ref{fig:quasar_lf}. We also assign host galaxies to the quasar population by matching the i-band magnitudes and redshifts of the quasars to those of the provided galaxy population\footnote{For the detailed process of quasar–host galaxy matching, please see \href{https://github.com/LSST-strong-lensing/slsim/blob/1d3609e30f6f047ef22cd2898a0845b550b19313/slsim/Sources/QuasarCatalog/README.md}{Quasar Host Matching}.}.

\subsection{Supernova sample}
\label{sec:supernovae}
A realistic supernova sample is required to simulate a realistic lensed supernova population. We sample supernovae from the supernova comoving density. The number density of type Ia supernovae is given by,
\begin{equation}
\label{eq:SNIa_n}
    n_{\rm SNIa} = \eta C_{\rm SNIa}\frac{\int_{0.1}^{t(z)} \rho_{\rm SFR}\left[z(t - t_D)\right]f(t_D) dt_D}{\int_{0.1}^{t(z=0)}f(t_D)dt_D},
\end{equation}
where $\eta$ and $C_{\rm SNIa}$ are constant factors with values of 0.04 and 0.032$M_\odot$, respectively \citep{BaldryandGlazebrook2003, Strigari_2005, Hopkins_2006, OguriMarshall2010}. $\rho_{\rm SFR}$ and $f(t_D)$ are the cosmic star formation rate and delay time distribution, respectively. These are given by,
\begin{equation}
\label{eq:rho_sfr}
    \rho_{\rm SFR} = \frac{(0.0118 + 0.08z)h}{1+(z/3.3)^{5.2}},
\end{equation}

and
\begin{equation}
\label{eq:delay_time}
    f(t_D) \propto t_D^{-1.08} (t_D > 0.1 \rm Gyr).
\end{equation}

Eqs. \ref{eq:rho_sfr} and \ref{eq:delay_time} are the best fit models to the observed data \citep{BaldryandGlazebrook2003, Totanietal2008}. The comparison of the number density computed from the simulated supernova sample with that of theoretical predictions from eq.\ \ref{eq:SNIa_n} is shown in Fig.\ \ref{fig:SNIa_n}.

To simulate supernova rates over a sky area and time, we integrate the supernova number density over a light-cone volume. To match our simulated supernova population with host galaxies, we first utilize the \texttt{SkyPyPipeline} class  to generate a host galaxy candidate catalog consisting of blue galaxies within the desired sky area. We select a host for each supernova by matching redshift values, then weighting the remaining host candidates according to a fitted empirical relation between supernova rate and strong star-forming host galaxy stellar mass, given by
\begin{equation}
\label{eq:Mstar_weight}
    \rho_{\rm SNIa} = n_{\rm mass}M_{\rm *},
\end{equation}
where $n_{\rm mass}$ is the slope of the best fit model, 0.74 ± 0.08, $M_{\rm *}$ is host galaxy stellar mass, and $\rho_{\rm SNIa}$ is supernova rate \citep{Sullivanetal2006}. The host galaxy is then selected randomly, with dependence on weight, from the redshift-matched host candidates. Although this work focuses on Type Ia supernovae, the \slsim\, framework can be easily extended to simulate core-collapse supernovae as well. Similar to the Type Ia case, functions for volumetric rates, luminosity distributions, and light curve modeling are available, and compatible \texttt{sncosmo} models exist, allowing users to implement core-collapse SN simulations with minimal additional effort.

We also assign realistic supernova offsets from the host galaxy center based on the empirical supernova offset ratio distribution \citep{Wangetal2013}. Fitting the empirical distribution to a log-normal probability density function yields parameters for shape, location, and scale equal to 0.764609, -0.0284546, and 0.450885 respectively, as shown in  Fig.\ \ref{fig:offset_ratio_dist}. Offset ratios are randomly drawn from this probability density function, transformed into offsets according to host angular size and random offset angles, converted into elliptical coordinates according to host ellipticity, and applied to each supernova.
 
\subsection{External shear and convergence}
\label{sec:shearconv}
In addition to multiple different deflector options, we also account for external shear and convergence in the lens model. One can choose to use a Gaussian mixture model to draw external shear and convergence for the deflector. In \slsim\,, the means, standard deviations, and weights of these models are obtained by fitting the data given in Table 2 of \cite{Schmidt:2023}. Then, we draw convergence and shear values randomly from this distribution and use them in the deflector mass model.

Another option for calculating external shear and convergence for a deflector is to use theoretical predictions for the line-of-sight mass structure. For this, we use a convergence map generated from the large-scale structure simulation in the \texttt{GLASS} package \citep{Tessoreetal2023}. This simulation provides a low-resolution but realistic model of the matter density field, capturing the contributions from cosmic structures along the line of sight. To account for local environmental effects around the deflector, based on the large-scale structure, we use the halos-rendering method, which is sampled from the halo mass function generated from the \texttt{colossus} package \citep{Diemer2018} and randomly placed in the line-of-sight light cone.

The halos located between the observer and the source or deflector introduce nonlinear effects on the convergence and shear. To account for this, we apply a correction factor to the total external convergence and shear, referred to as the non-linear correction:
\begin{equation}
\label{eq:correction kappa}
    1 - \kappa_{\text{ext}} = \frac{(1-\kappa_{\text{OD}})(1-\kappa_{\text{OS}})}{1-\kappa_{\text{DS}}}
\end{equation}

\begin{equation}
\label{eq:correction convergence}
    \gamma_{\text{ext}} = \sqrt{(\gamma_{\text{OD1}} + \gamma_{\text{OS1}} - \gamma_{\text{DS1}})^2 + (\gamma_{\text{OD2}} + \gamma_{\text{OS2}} - \gamma_{\text{DS2}})^2}
\end{equation}

Shown in Equation~\ref{eq:correction kappa} and Equation~\ref{eq:correction convergence}, total external convergence (\( \kappa_{\text{ext}} \))  is influenced by convergence from the halos between the observer and the deflector (\( \kappa_{\text{OD}} \)), the observer and the source (\( \kappa_{\text{OS}} \)), and the deflector and the source (\( \kappa_{\text{DS}} \)). Total external shear (\( \gamma_{\text{ext}} \)) is computed by observer-deflector (\( \gamma_{\text{OD}} \)), observer-source (\( \gamma_{\text{OS}} \)), and deflector-source (\( \gamma_{\text{DS}} \)). The components of shear in two orthogonal directions are represented as \(\gamma_1\) and \(\gamma_2\). 

By integrating the large-scale structure with these randomly positioned halos, we generate a combined distribution of external convergence and shear that reflects the deflector’s specific redshift. Once the distributions are constructed, we sample pairs of external shear and convergence values, which are then incorporated into our lensing model. A comprehensive explanation of the methodology, including detailed discussions on the generation of external convergence-shear maps, the sampling of the halo mass function, and the selection effects introduced by the line-of-sight structure on galaxy-galaxy lensing, is presented in \citep{Tangetal2025}.

To customize the external shear and external convergence configuration, users can make use of the \texttt{los\textunderscore config} or \texttt{los\textunderscore dict} parameters in the \texttt{LensPop} or \texttt{Lens} classes. The \texttt{los\textunderscore config} parameter accepts an instance of the \texttt{LOSConfig} class, whereas \texttt{los\textunderscore dict} allows for the direct specification of parameters to configure the \texttt{LOSConfig} class. Users have the option to include or exclude external convergence and shear in the realizations. If these effects are included, users can choose between a Gaussian mixture fit or theoretical predictions to model line-of-sight effects. For theoretical predictions, users can optionally apply non-linear corrections. By default, the package chooses the theoretical predictions distribution without the corrections.

To enable non-linear corrections, users must first provide a nonlinear correction file via the \texttt{nonlinear\textunderscore correction\textunderscore path} parameter in \texttt{LOSConfig} class, as the package does not include this distribution by default. Additionally, users can opt for a simpler configuration by directly inputting external convergence and shear values for specific realizations or other purposes. For more details on the available distributions, visit the public link. \footnote{The Line of Sight External Convergence and Shear Distributions Files: \href{https://github.com/LSST-strong-lensing/data_public/tree/main/Line_of_sight_kg_distributions}{https://github.com/LSST-strong-lensing/data\_public/tree/main/Line\_of\_sight\_kg\_distributions}}

\subsection{Velocity Dispersion module}
\label{sec: vel_disp}
This module implements a suite of stellar velocity dispersion models for lens galaxy populations. It provides routines to compute luminosity-weighted and aperture-averaged stellar velocity dispersions based on a range of physical models, including composite mass profiles (Hernquist + NFW), NFW halos, and elliptical power-law profiles. In addition, the module includes a velocity dispersion function, calibrated to SDSS data \citep{bernadietal2010, Choietal2007}, which is used to generate statistically representative samples of galaxies over a redshift range. This follows an abundance-matching method that assigns velocity dispersions to galaxies based on stellar mass rankings. These capabilities allow for realistic modeling of deflector kinematics across a wide variety of galaxy populations and redshifts.

Furthermore, we implement a light-to-mass model that uses the light profile of deflector galaxies to calculate their central stellar velocity dispersion. First, it computes the luminosity of each deflector galaxy using k-corrected {\it g}, {\it r}, and {\it i} band magnitudes along with photometric redshifts. The model assumes evolution of the galaxy luminosity, in which the characteristic magnitude $M^{r}_{*}$ brightens by 1.5 magnitudes from $z = 1$ to $z = 0$, based on results from the DEEP2 and COMBO-17 surveys \citep{Belletal2004}. Next, it applies the $L$–$\sigma$ scaling relation, derived for elliptical galaxies using spectroscopic measurements in \cite{Choietal2007} (the default choice), to estimate the stellar velocity dispersion of the deflector:
\begin{equation}
\label{eq:L_sigma_relation}
\sigma_{v} = 161 \left( \frac{L}{L_{*}} \right)^{\frac{1}{2.32}} , \text{km} , \text{s}^{-1}.
\end{equation}
Alternatively, the scaling relation derived from the weak-lensing measurements in \cite{Parkeretal2007} is also implemented and can be used in place of eq.\ \ref{eq:L_sigma_relation}.

\subsection{Microlensing}
\label{sec: microl}
The \slsim\ software package incorporates the ability to simulate lightcurves, including the effects of microlensing induced by the stellar population of the deflector galaxy. The generation of microlensing lightcurves primarily necessitates information regarding the source morphology and the deflector's lensing properties---specifically, convergence ($\kappa$), shear ($\gamma$), and stellar convergence ($\kappa_*$)---at the image positions \cite{Vernardos:2024}. Currently, the module supports two source morphologies:
\begin{itemize}
    \item A \textbf{Gaussian Source}, implemented via the \texttt{GaussianSourceMorphology} class.
    \item A \textbf{Static Accretion Disk AGN Source}, available through the \texttt{AGNSourceMorphology} class.
\end{itemize}
Future development includes the planned implementation of a supernova morphology class and variable source morphologies. Furthermore, the module provides functionality to generate magnification maps using the \texttt{MagnificationMap} class, provided that the deflector properties are available at the specified image positions.

\begin{figure}
    \centering
    \includegraphics[width=\linewidth]{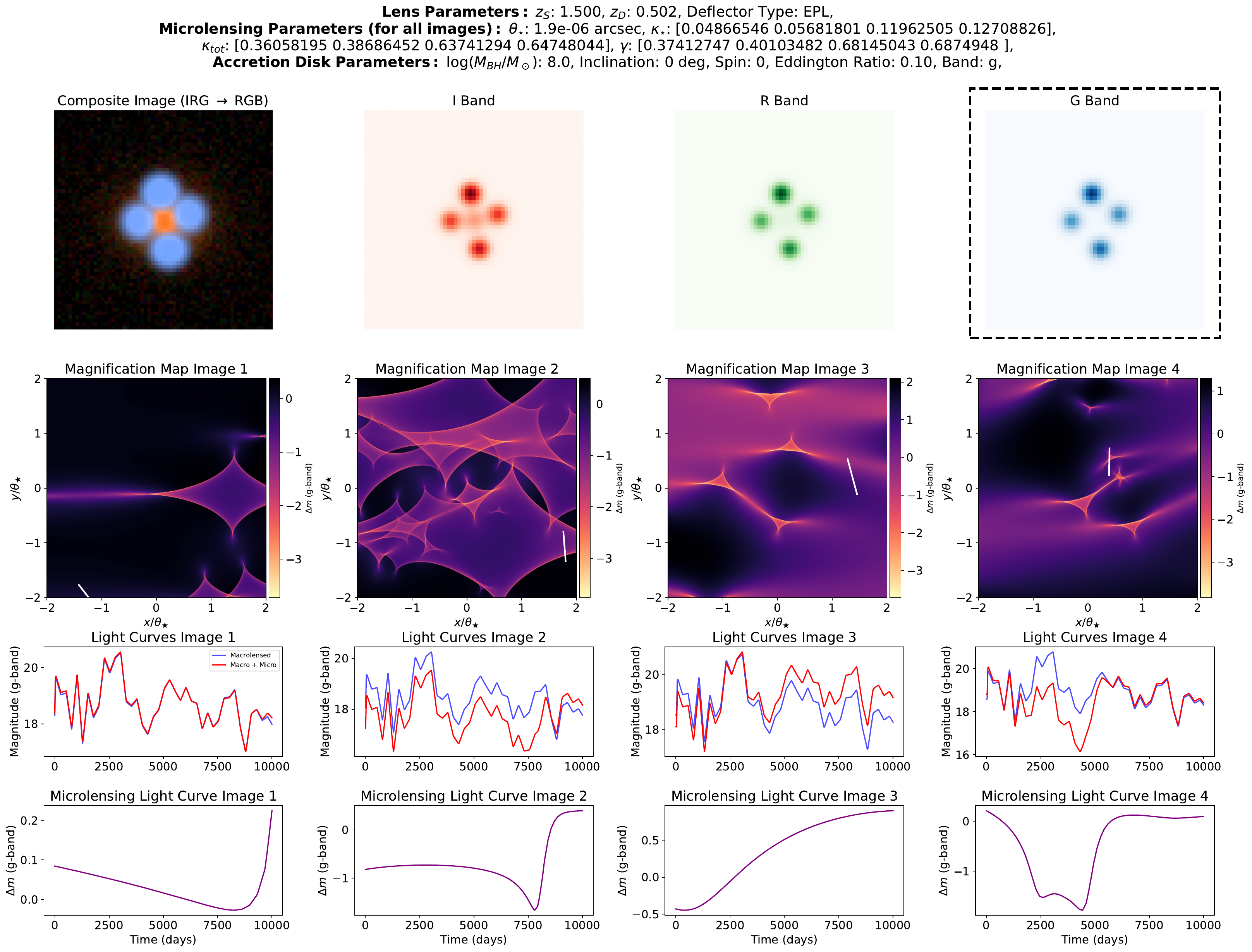}
    \caption{An example of a simulated, time-variable, quadruply lensed quasar generated with SLSim. \textit{Top Row:} A composite RGB color image created from the individual I, R, and G bands. The G-band image (outlined by the dashed box) is used for the analysis shown in the lower panels. \textit{Second Row:} Microlensing magnification maps for each of the four lensed images, revealing the complex caustic networks. \textit{Third Row:} Simulated g-band light curves. The red curves show the total brightness, which includes intrinsic quasar variability, macrolensing, and microlensing effects. For comparison, the light curves without the microlensing component are shown in blue. \textit{Bottom Row:} The isolated microlensing contribution, plotted as the deviation in magnitude ($\Delta m$) from the macro-model.}
    \label{fig:slsim_microlensing_example}
\end{figure}

Within the \texttt{Lens} class in \slsim\, one can generate lightcurves for variable sources with the inclusion of microlensing by using the \texttt{point\_source\_magnitude} method. One such example of a quadruply lensed variable quasar source is shown in Fig.~\ref{fig:slsim_microlensing_example}.

\subsection{Back-end packages}
\label{sec: backends}
In this first release of \slsim\, we use different packages for the astronomical calculation and mock data simulation. These packages are listed below:
\begin{itemize}
    \item \texttt{lenstronomy}\footnote{\href{https://github.com/lenstronomy/lenstronomy}{https://github.com/lenstronomy/lenstronomy}}: This is a multi-purpose software package to model strong gravitational lenses \citep{lenstronomy, lenstronomyII}. We use lenstronomy for many strong gravitational lensing related calculations. \texttt{lenstronomy} also provides an image simulation API and we use this API in \slsim\,.
    \item \texttt{skypy}\footnote{\href{https://github.com/skypyproject/skypy}{https://github.com/skypyproject/skypy}}: This is a software package that simulates the populations of astronomical objects \citep{skypyetal2021}. We use \texttt{skypy} to simulate the galaxy population. 
    \item \texttt{sncosmo}\footnote{\href{https://github.com/sncosmo/sncosmo}{https://github.com/sncosmo/sncosmo}}: This is a package for supernova cosmology analysis \citep{Barbaryetal2016}. In \slsim\,, we use \texttt{sncosmo} to generate supernova light curves for the supernova sample.
    \item \texttt{colossus}\footnote{\href{https://bdiemer.bitbucket.io/colossus/}{https://bdiemer.bitbucket.io/colossus/}}: This is a software package that deals with the large-scale structure in the universe \citep{Diemer2018}. We use the halo-mass function from colossus to sample halos.
    \item \texttt{microlensing}\footnote{\href{https://github.com/weisluke/microlensing.git}{https://github.com/weisluke/microlensing}}: This software package deals with the generation of microlensing magnification maps on GPU.
\end{itemize}
In addition to the packages mentioned above for astronomical calculations, \slsim\, depends on the following standard \texttt{Python} libraries: Numpy \citep{Harris_et_al2020}, SciPy \citep{Virtanen_et_al2020}, Matplotlib \citep{Hunter2007}, and Astropy \citep{Astropy}.

\section{The LSST science pipeline module}
\label{sec:lsst_science}
One of the goals of \slsim\, is to produce realistic strong lens images for the LSST survey. To achieve this, we need the noise properties, PSF, and zero-point magnitude from the LSST observations to generate accurate lens images. We have two options for accessing the simulated LSST observation properties, which are described below.

\subsection{DP0 data}
\label{sec:dp0}
Currently, we have simulated data for LSST as part of Data Preview zero (DP0), which includes realistic observational properties. We use DP0.2 data which is the second version of the data preview\footnote{Documentation for the DP0.2 data can be found here: \href{https://dp0-2.lsst.io}{https://dp0-2.lsst.io}}. The DP0.2 data are the 300 $deg^2$ of simulated, LSST-like exposures, imaging products, and catalogs generated by the DESC for their Data Challenge 2 (DC2) \citep{DC2_paper}. Injecting our simulated lenses into the DP0.2 data provides us with the realistic lens image catalog, allowing us to test source injections and facilitate the transition to real data.

In \slsim\,, we have a dedicated \texttt{lsst\_science\_pipeline} module. Primarily, this module performs the following operations: 1) Queries DP0.2 data/images. 2) Produces DP0.2 image cutouts. These cutouts are small excerpts of large-size images with specified pixel sizes. Additionally, it extracts all the observational properties of these cutouts such as exposure map, PSF, and zero-point magnitudes. These properties can be passed on to the image simulation module while generating the lens image. 3) Injecting simulated lens images into the corresponding DP0.2 image cutouts. The \texttt{lsst\_science\_pipeline} is connected with the other modules of \slsim\, such as image simulation module and the \texttt{LensPop} class, and is shown in Fig.\ \ref{fig:illustration} in green color. The \texttt{lsst\_science\_pipeline} module makes the best use of LSST software packages developed by the LSST project team for many purposes. Querying LSST project data products requires access to LSST servers. One of the key properties of the \texttt{lsst\_science\_pipeline} module is that it integrates \slsim\, with the Rubin Science Platform (RSP)\footnote{The Rubin Science Platform is part of the Vera C. Rubin Observatory, facilitating access, analysis, and collaboration on the vast datasets from LSST survey. It includes APIs, a portal, and a Jupyter notebook in integrated form.} and is available through it.

The \texttt{lsst\_science\_pipeline} module primarily utilizes the LSST Butler\footnote{Butler is an LSST Data Access framework, and its software can be found at \href{https://github.com/lsst/daf_butler}{https://github.com/lsst/daf\_butler}} \citep{lsst_butler},  a data access and management system, to access DP0 data. In some cases, it also uses the Table Access Protocol (TAP) query service to access metadata for sets of DP0 single-exposure images. After querying the DP0 image of a patch and tract, the module creates cutouts from this image. For static cutout images, it queries deepCoadd images, and for time series cutouts, it queries single-exposure images. Subsequently, it selects a random lens (or a user can provide a specific lens) from a given \texttt{LensPop} instance and injects it into the prepared DP0 cutout image. This process can be repeated for multiple lenses and cutout images. Users specify the number of injected lenses they want to achieve, and this module automates the process. Finally, it returns an Astropy Table containing injected strong lenses with the necessary metadata. A sample of lenses injected into the DP0 cutouts is shown in Figs.\, \ref{fig:quasar_lc} and \ref{fig:injected_sample}.

\subsection{Rubin Operations Simulator (OpSim)}
\label{sec:opsim}
Injecting lenses to the DP0 data is a very slow process, particularly for time-variable lenses. To inject time-series images of variable lenses into the DP0.2 data, we need to produce time-series cutouts of single-visit images of the DP0.2 data. This includes querying all single-visit images at the coordinate of interest. This process takes $\sim 30-35$ seconds for a single coordinate. Therefore, to speed up this process, we use an alternative approach i.e.\ the Rubin Operations Simulator (OpSim). 

OpSim simulates the field selection and image acquisition process for LSST throughout the planned 10-year survey period \citep{Delgado2014, Delgado2016, Naghib2019}. Information on each simulated telescope pointing is recorded in an output data product, organized as an `sqlite' database, with each pointing represented as a row in the database. For our purposes, we need to query all observations in the database and find the single visits and metadata which correspond to a particular coordinate point. To do this efficiently, \cite{Biswasetal2020} developed OpSimSummary\footnote{\href{https://github.com/LSSTDESC/OpSimSummary}{https://github.com/LSSTDESC/OpSimSummary}} , which uses a Tree data structure to speed up the calculations and the sorting process. To comply with the newest OpSim database structure, OpSimSummaryV2\footnote{\href{https://github.com/LSSTDESC/OpSimSummaryV2}{https://github.com/LSSTDESC/OpSimSummaryV2}}  was developed by Bastien Carreres, which is what we use in SLSim to access the observations corresponding to specific positions in the sky. We query the metadata describing each visit and use the observation time, filter, zero-point magnitude, PSF FWHM, exposure time, and sky brightness to simulate single-visit backgrounds. These backgrounds can be used as time series images into which simulated lens images can be injected. This procedure allows us to recover multiband observations from OpSim corresponding to a long list of sky coordinates in around 40 seconds.

\section{Roman Space Telescope Simulation}
\begin{figure}
\centering
    \includegraphics[width=\linewidth]{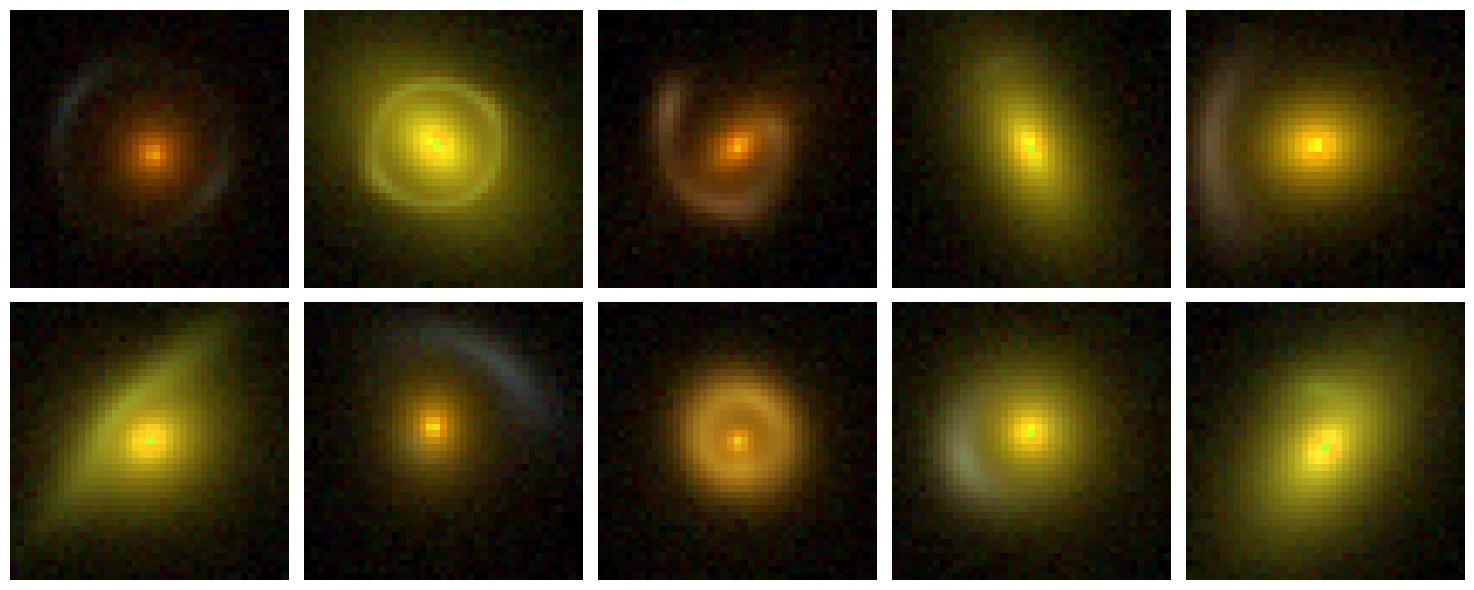}\par
    \hspace{10mm}
    \includegraphics[width=\linewidth]{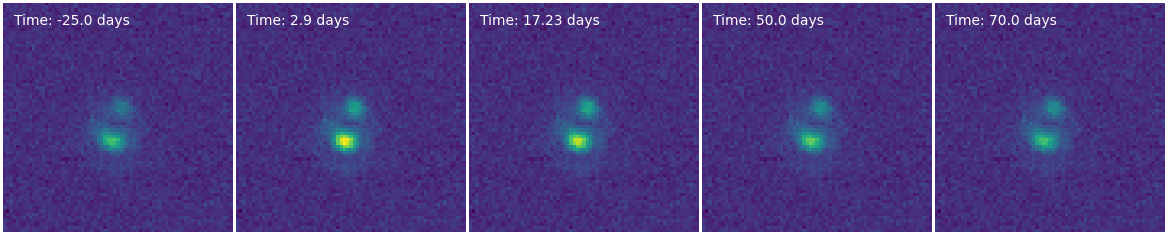}\par
\caption{Roman static and variable lens images simulated using \slsim\,. Top panel: Examples of galaxy-galaxy lenses in RGB color. Bottom panel: Example of time-series images of a lensed supernova in the F129 band.}
\label{fig:Romanlens}
\end{figure}

We have extended the \slsim\, simulation pipeline to generate realistic strong lensing mock data tailored for the Nancy Grace Roman Space Telescope. The pipeline simulates high-resolution galaxy-galaxy strong lenses and time-domain sequences of lensed supernova images, incorporating Roman-specific parameters such as instrument resolution, psf, and noise properties. \slsim\, obtains the Roman instrument properties from two main sources: \texttt{GalSim} package \citep{galsim2015}, which provides detector effects based on the Wide Field Instrument (WFI) technical data\footnote{More information on Roman instrumentation can be found at: \href{https://roman.gsfc.nasa.gov/science/WFI_technical.html}{https://roman.gsfc.nasa.gov/science/WFI\_technical.html}}, and \texttt{STPSF} package \citep{webpsf2014}, which provides band- and detector-dependent psfs. Samples of roman image simulations are shown in Fig.\ \ref{fig:Romanlens}. The \slsim, package has already been used in the simulation of the Roman view of strong gravitational lenses, with further details provided in \cite{Wedigetal2025}. These simulations provide training and validation data for lens discovery algorithms, lens modeling techniques, and cosmological inference pipelines. Therefore, \slsim\, enables robust end-to-end testing of strong lensing workflows in preparation for future space-based surveys.

\section{Validation with DESC resources}
\label{sec:validation}
\slsim\, has routines for the simulation of lens populations, image simulation, and lens injection into DP0 data. In this section, we present validations that go beyond unit testing for full system integration. We validate different aspects of \slsim\, by comparing its results with existing DESC resources and products. Specifically, we validate the image simulation in \slsim\,. For this purpose, we select a galaxy from the DP0.2 dataset and extract all relevant catalog values such as magnitude, shear, convergence, ellipticity, angular size, and Sersic indices. Furthermore, we retrieve the noise properties, PSF, and zero-point magnitude from the DP0.2 data. Using this information, we generate images of galaxies with \slsim\,'s image simulation module and compare it to the corresponding DP0.2 image of the galaxy for validation (see Fig.\ \ref{fig:image_validation}\footnote{The notebook performing this image comparison can be found here: \href{https://github.com/LSST-strong-lensing/slsim/blob/ec7335968fc20d74fe805db2aa4aa3093bb983c0/notebooks/validation_notebooks/image_validation.ipynb}{Image validation notebook}.}).

\begin{figure*}
    \includegraphics[width=\linewidth]{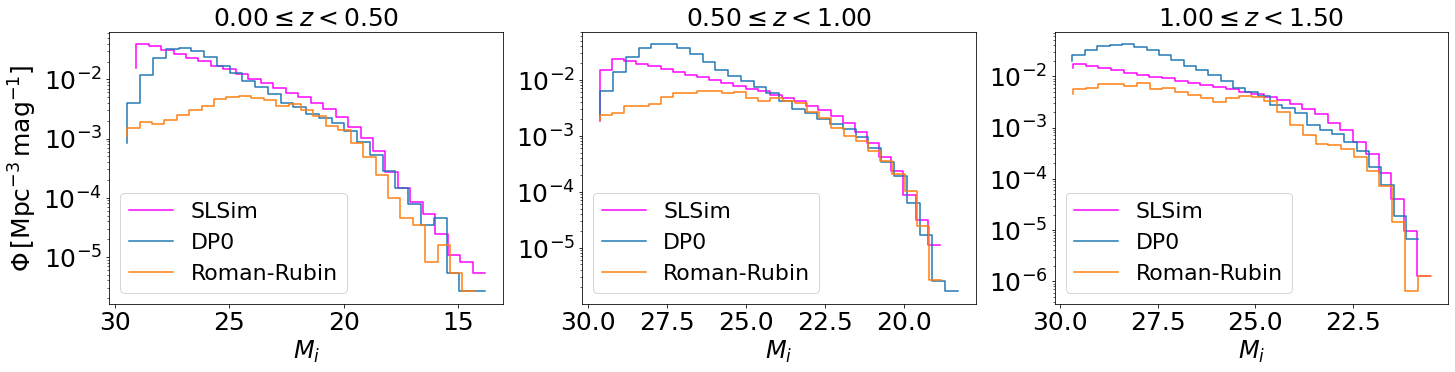}\par
\caption{Comparison between the luminosity function of \slsim\, generated galaxy population with the luminosity function of DP0 galaxies \citep{DC2_paper} and Roman-Rubin galaxies \citep{RomanRubinsim2023}. The luminosity function for each galaxy population is computed in the i-band magnitude. The left, middle, and right plots compare the luminosity functions in the redshift ranges $0.00 \leq z < 0.50$, $0.50 \leq z < 1.00$, and $1.00 \leq z < 1.5$, respectively.
} 
\label{fig:slsim_dp0_galaxy_distribution}
\end{figure*}

One of the key components and a prerequisite to simulate accurate populations of strong lenses is to have accurate descriptions of the population of deflector and source galaxies in numbers, magnitudes and redshift. We compare the Schechter luminosity function of elliptical galaxies generated using \slsim\, with the SDSS bestfit model predictions \citep{Monteroetal2009} in different redshift bins. The comparison is shown in Fig.\ \ref{fig:SF_comp}. In this comparison, we examine two components of the Schechter function. The first component represents the dwarf elliptical galaxy population, while the second component describes the massive elliptical galaxy population.\footnote{The detailed description of this plots and all the required parameters are given in the notebook: \href{https://github.com/LSST-strong-lensing/slsim/blob/ec7335968fc20d74fe805db2aa4aa3093bb983c0/notebooks/validation_notebooks/slsim_dc2_diffsky_galaxy_distribution_comparision.ipynb}{Notebook for the validation of galaxy distribution}.}  We demonstrate that our galaxy populations are correctly drawn from the observed galaxy luminosity function. We also compare the luminosity function of all galaxies generated using \slsim\, with the luminosity function of the DP0 galaxy population and the galaxy population of the Roman-Rubin simulation \citep{RomanRubinsim2023}. \footnote{Roman-Rubin galaxy catalog is the simulated galaxy catalog for joint Roman and Rubin observations.} The comparison is shown in Fig.~\ref{fig:slsim_dp0_galaxy_distribution}. From this figure, we can see that the luminosity function of \slsim\, generated galaxies in different redshift bins matches the luminosity function of DP0 and Roman-Rubin simulated galaxies very well in the bright end; however, at the faint end, all three populations differ from each other. We can safely ignore these differences in the simulation because these faint galaxies are very difficult to observe.

\section{Notebook suite}
\label{sec:notebook}
In addition to comprehensive documentation, our software package includes a notebook suite that features a series of example notebooks demonstrating the core functionality of \slsim\, in action. Each notebook includes detailed explanations, step-by-step instructions, and illustrative examples that showcase how to implement and adapt the code for different use cases.\footnote{All these notebooks can be found here: \href{https://github.com/LSST-strong-lensing/slsim/tree/main/notebooks}{\slsim\, notebook suite}} These notebooks could promote collaboration and knowledge sharing within the user community.

\section{Future Directions}
\label{sec:future}
Currently, the package is designed to support and prepare for various aspects of strong lensing science within LSST-DESC and the Strong Lensing Science Collaboration. With several projects underway, \slsim\, assists by generating simulated data for tasks such as creating machine learning training sets for lens searches and search challenges in both the static and transient domain. Additionally, \slsim\, will be able to quantify lensing selections and will be an essential tool to assess end-to-end lens search and analysis frameworks. In particular, the realistic lens images generated by \slsim\, can be used to test SLSC-DESC strong lensing analysis pipelines, identify potential biases, and quantify their impact on the inference of cosmological parameters. Large and realistic training sets can also be used in lens modeling algorithms. \slsim\, can assist in any future lens modeling challenge based on machine learning methods. We also plan to use \slsim\, for the forecast of rare and exotic lensing events, which could help optimize strategies for their detection.

Running the LSST data stream through a Difference Image Analysis (DIA) pipeline is one of the crucial aspects of finding lensed transients from the survey data. Validating and understanding the performance of a DIA pipeline with regard to specific transients is important for detecting the transients in real time. We envision using \slsim\, to test the performance of the LSST-DESC DIA pipeline using both resolved and unresolved lensed transient images simulated using \slsim\,. The extraction of pixel-to-light curve data of multiply imaged quasars and transients and subsequently converting these light curves to time-delay measurements are vital for conducting robust time-delay cosmography. 

The growing user and developer base of \slsim\, ensures continuous improvements to the software, and the addition of relevant features as LSST data is anticipated in the near futures and the science applications are growing. An ultimate goal is to achieve end-to-end cosmological inference, which includes lens searches, pipeline validations, and analysis to derive accurate cosmological constraints. Through collaborative community efforts, our aim is to advance \slsim\,’s capabilities as a key support role in these upcoming science analyses. In the future, we envision extending \slsim\ to support simulations tailored for other major surveys such as Roman (already supported) and Euclid, enabling cross-survey synergy and preparing for multi-facility strong lensing pipelines in the LSST era.

\section{Conclusion}
\label{sec:conclusion}
In this paper, we introduce \slsim\,, a comprehensive and user-friendly package designed to simulate realistic strong lensing populations and their images, tailored specifically for the Vera C. Rubin Observatory but adaptable to other observatories such as Roman and Euclid. We detailed the core components, functionalities and  features of \slsim\,, emphasizing its capabilities in simulating both static and variable lens cases. Our validation tests demonstrated the package's accuracy and performance, showcasing its versatility in simulating a wide range of strong lensing phenomena. As we look towards the future, \slsim\, stands as a robust tool ready to facilitate the efficient identification and modeling of strong gravitational lenses, thus significantly advancing our capacity to explore the rich scientific potential of forthcoming large-scale surveys in the domain of strong lensing.

\begin{appendix}
\section{Verification of quasar and supernovae samples.}
\label{sec:appendix}
\begin{figure*}
    \includegraphics[width=\linewidth]{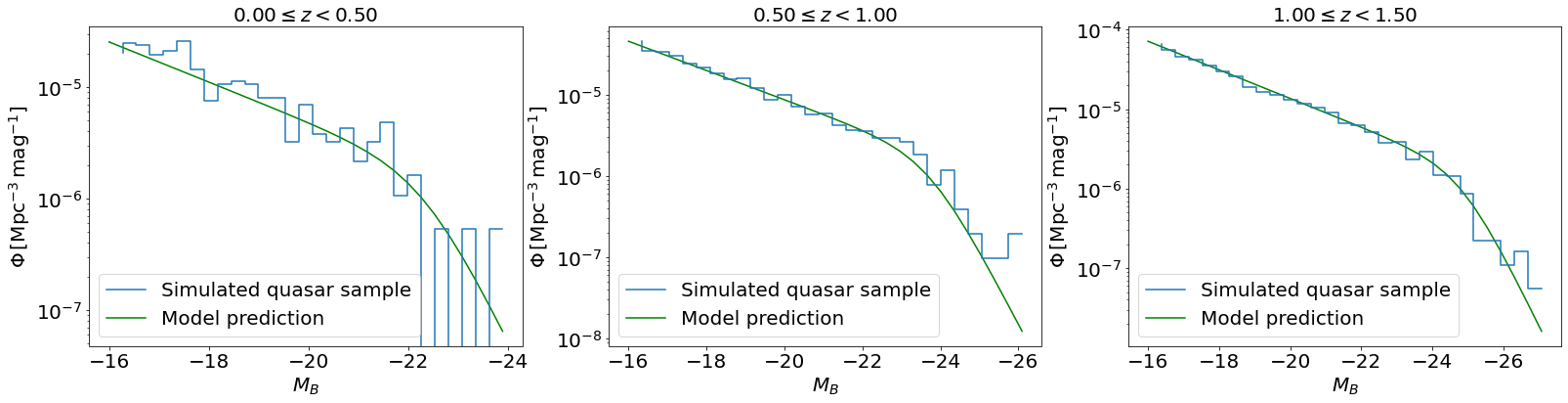}\par
\caption{Comparison of the quasar luminosity function obtained from the simulated quasar sample with that derived from the same model in different redshift bins. The luminosity function is computed with the absolute magnitude. The plots on the left, middle and right compare the luminosity functions in the redshift ranges $0.00 \leq z < 0.50$, $0.50 \leq z < 1.00$, and $1.00 \leq z < 1.5$, respectively. This comparison verifies that the sampling of quasars from the luminosity function is performed correctly.}
\label{fig:quasar_lf}
\end{figure*}

\begin{figure}
    \includegraphics[width=\linewidth]{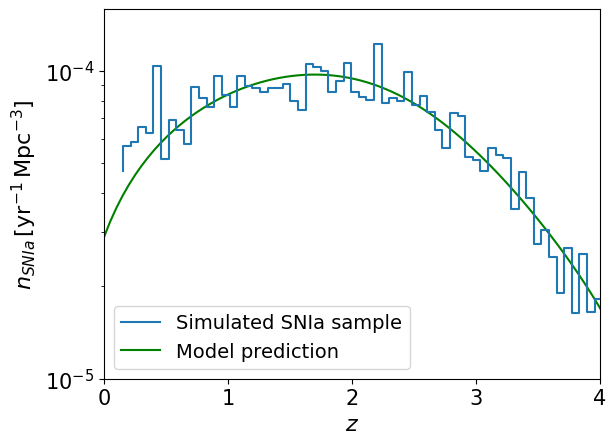}\par
\caption{Comparison of the supernovae number density obtained from the simulated supernovae sample with that derived from the same model \citep{OguriMarshall2010}. This comparison verifies that the sampling of supernova redshifts using number density is performed correctly.}
\label{fig:SNIa_n}
\end{figure}

Unit testing individual routines in \slsim, is one thing. System integration testing to guarantee the integration and resulting population-level generated samples, is another crucial part in the validation process.
For the simulation of a realistic quasar sample within a specified sky area, we sample quasars from the quasar luminosity function given in Equation \ref{eq:qlf}. To verify our sampling method, we compare the theoretical quasar luminosity function with the sample-derived luminosity function in Fig.\ \ref{fig:quasar_lf}. In this figure, one can see that our sample of quasars in different redshift bins agrees with the theoretical predictions. Similarly, for the simulation of a realistic Type Ia supernova sample within the specified sky area, we sample supernovae from the Type Ia supernovae rate given in eq.\ \ref{eq:SNIa_n}. In Figure \ref{fig:SNIa_n}, we compare the theoretical supernova rate with the simulated sample-derived rate. In this figure, one can see that our supernovae samples agree with the theoretical predictions.
\end{appendix}

\acknowledgments
This paper has undergone internal review in the LSST Dark Energy Science Collaboration. The internal reviewers were Richard Kessler and Justin R. Pierel.

The \slsim\ package was led and developed by Narayan Khadka, who designed the framework, implemented and validated the majority of its components, and wrote the manuscript. Simon Birrer initiated the overall code design and provided continuous mentorship to the developer team. Key contributions to major modules were made by Henry Best, who developed and tested the AGN variability module and prepared example use cases, and Paras Sharma, who contributed to both the microlensing and quasar modules. Halo-based deflector models were developed by Katsuya T. Abe and Masamune Oguri, while Xianzhe Tang implemented the external shear and convergence components and contributed to the manuscript. Felipe Urcelay developed the group- and cluster-scale deflector module and contributed to writing the paper. Carly Mistick implemented the supernova population and supernova–host galaxy connection, and Emrecan M. Sonmez developed the quasar population module.
Additional major components include the integration of OpSim by Nikki Arendse, the Roman image simulation module by Alan Huang, light-to-mass modeling by Vibhore Negi and Anupreeta More, the supernova class by Justin R. Pierel, the Euclid configuration by Yixuan Shao, and the implementation of real galaxy images in lens simulations by Rahul Karthik.
Additional contributions to the development, discussion, and improvement of SLSim were made by Aadya Agrawal, Padma Venkatraman, Phil Holloway, Bryce Wedig, Jacob O. Hjortlund, Tian Li, Sydney Erickson, Anowar Shajib, Bruno Sánchez, Timo Anguita, Pedro Bessa, Clecio R. Bom, Sofia Castillo, Thomas Collett, Tansu Daylan, Steven Dillmann, Margherita Grespan, Erin E. Hayes, Rémy Joseph, Richard Kessler, Phil Marshall, Veronica Motta, Gautham Narayan, Matt O’Dowd, Aprajita Verma, and Giorgos Vernardos.

The DESC acknowledges ongoing support from the Institut National de 
Physique Nucl\'eaire et de Physique des Particules in France; the 
Science \& Technology Facilities Council in the United Kingdom; and the
Department of Energy and the LSST Discovery Alliance
in the United States.  DESC uses resources of the IN2P3 
Computing Center (CC-IN2P3--Lyon/Villeurbanne - France) funded by the 
Centre National de la Recherche Scientifique; the National Energy 
Research Scientific Computing Center, a DOE Office of Science User 
Facility supported by the Office of Science of the U.S.\ Department of
Energy under Contract No.\ DE-AC02-05CH11231; STFC DiRAC HPC Facilities, 
funded by UK BEIS National E-infrastructure capital grants; and the UK 
particle physics grid, supported by the GridPP Collaboration.  This 
work was performed in part under DOE Contract DE-AC02-76SF00515.

Narayan Khadka is supported by the DOE/DESC, Stony Brook University, and the Schmidt Futures Foundation. Henry Best acknowledges support from the GA\v{C}R Junior Star grant GM24-10599M. Support for Matt O'Dowd was provided by Schmidt Sciences, LLC. Timo Anguita acknowledges support from ANID-FONDECYT Regular Project 1240105, ANID Millennium Science Initiative AIM23-0001, and the ANID BASAL project FB210003. Tansu Daylan acknowledges support from the National Aeronautics and Space Administration (NASA) under grant number 80NSSC24K0095 issued by the Astrophysics Division of the Science Mission Directorate, as well as support from the McDonnell Center for the Space Sciences at Washington University in St. Louis. Bryce Wedig acknowledges support from a NASA Future Investigators in NASA Earth and Space Science and Technology (FINESST) research grant (Grant No. 80NSSC24K1481). Erin E. Hayes is supported by a Gates Cambridge Scholarship (OPP1144). V. Motta acknowledges partial support from Fondecyt Regular 1231418 and Centro de Astrofísica de Valparaíso CIDI 21.


\bibliographystyle{plain}
\bibliography{biblio}



\end{document}